\documentclass[aps,prd,amsmath,amsfonts,amssymb,eqsecnum,nofootinbib,
  floatfix,secnumarabic,preprintnumbers,superscriptaddress,nobalancelastpage,
  onecolumn,notitlepage,biblatex]{revtex4-1}
\usepackage[utf8]{inputenc}
\usepackage{color,graphicx,hyperref,dsfont,slashed,multirow,subfig}
\usepackage{dsfont}
\usepackage{dcolumn}   
\usepackage{bm}        
\usepackage[english]{babel}
\usepackage{adjustbox}
\usepackage{pifont}
\usepackage{multirow}
\newcommand{\met}{\ensuremath{{\not\mathrel{E}}_T}}

\newcommand{\cmark}{\ding{51}}%
\newcommand{\xmark}{\ding{55}}%

\def\a{\alpha}

\def\r{\rho}
\def\s{\sigma}
\def\t{\tau}
\def\m{\mu}
\def\n{\nu}
\def\k{\kappa}
\def\th{\theta}
\def\g{\gamma}\def\G{\Gamma}
\def\L{\Lambda}\def\l{V}
\def\D{\Delta}
\def\la{\langle}
\def\ra{\rangle}
\def\o{\omega}\def\O{\Omega}
\def\d{\delta}
\def\p{\partial}

\def\oxthree{{\cal O}(x^3) }

\def\h{h^0}

\def\hpm{H^{\pm}}
\def\hmp{H^{\mp}}
\def\wpm{W^{\pm}}
\def\wmp{W^{\mp}}

\def\half{\textstyle{\frac{1}{2}}}

\def\bdoc{\begin{document}}
\def\edoc{\end{document}}
\def\be{\begin{equation}}
\def\ee{\end{equation}}

\def\beq{\begin{eqnarray}}
\def\eeq{\end{eqnarray}}
\def\ben{\begin{enumerate}}
\def\een{\end{enumerate}}
\def\la{\langle}
\def\ra{\rangle}
\def\a{\alpha}
\def\g{\gamma}\def\G{\Gamma}
\def\d{\delta}\def\D{\Delta}
\def\e{\epsilon}
\def\z{\zeta}

\def\th{\theta}
\def\k{\kappa}
\def\l{\lambda}
\def\m{\mu}
\def\n{\nu}
\def\o{\omega}
\def\p{\pi}
\def\r{\rho}
\def\s{\sigma}
\def\t{\tau}
\def\L{{\Lambda}}
\def\S{\Sigma }
\def\gsim{\; \raisebox{-.8ex}{$\stackrel{\textstyle >}{\sim}$}\;}
\def\lsim{\; \raisebox{-.8ex}{$\stackrel{\textstyle <}{\sim}$}\;}
\def\gtrsim{\gsim}
\def\lessim{\lsim}
\def\loc{{\rm local}}
\def\vm{v_{\rm max}}
\def\bh{\bar{h}}
\def\del{\partial}
\def\nab{\nabla}
\def\half{{\textstyle{\frac{1}{2}}}}
\def\fourth{{\textstyle{\frac{1}{4}}}}
\def\h{\mathscr{H}}
\def\bD{{\bf D}}
\def\bE{{\bf E}}
\def\bF{{\bf F}}
\def\bB{{\bf B}}
\def\bP{{\bf P}}
\def\bV{{\bf v}}
\def\bv{{\bf v}}
\def\bx{{\bf x}}
\def\by{{\bf y}}
\def\bz{{\bf z}}
\def\ba{{\bf a}}
\def\bd{{\bf d}}
\def\bs{{\bf s}}
\def\bn{{\bf n}}
\def\bp{{\bf p}}

\def\O{\Omega}

\def\br{{\bf r}}
\def\bnab{{\bf \nab}}

\def\tE{\tilde{E}}
\def\tL{\tilde{L}}
\def\Horava{Ho\v{r}ava }

\def\oxtwo{\mathscr{O}\left(x^2\right)}
\def\oxthree{\mathscr{O}\left(x^3\right)}
\def\oxfour{\mathscr{O}\left(x^4\right)}
\def\oxfive{\mathscr{O}\left(x^5\right)}

\def\ph{\phantom}

\begin{document}

\title{Charged Higgs Discovery Prospects}

\author{Baradhwaj Coleppa}
\email {baradhwaj@iitgn.ac.in}
\affiliation {Physics Discipline,
Indian Institute of Technology-Gandhinagar, Palaj Campus, Gujarat 382355, India}


\author{Agnivo Sarkar}
\email{agnivo.sarkar@iitgn.ac.in}
\affiliation {Physics Discipline,
Indian Institute of Technology-Gandhinagar, Palaj Campus, Gujarat 382355, India}

\author{Santosh Kumar Rai}
\email{skrai@hri.res.in}
\affiliation{Regional Centre for Accelerator-based Particle Physics,
Harish-Chandra Research Institute, HBNI,
Chhatnag Road, Jhusi, Prayagraj (Allahabad) 211019, India}

\date{\today}
\begin{abstract}
We study the discovery prospects of the charged Higgs boson in the context of multi Higgs models in certain BSM scenarios. We classify models into three categories based on the charged Higgs coupling properties: gaugophobic, fermiophobic, and chromophobic. In each case, we identify viable modes of discovery, and present LHC analysis for discovery. We find that extensions of the Standard Model in which the charged Higgs does not couple to colored particles offer the best possible avenues for discovery.
\end{abstract}

\maketitle

\section{Introduction}
The Standard Model (SM) \cite{Weinberg:1967tq} of particle physics has been a phenomenal success in describing three of the four fundamental forces and myriad experiments have now firmly established its particle content and to a large extent, the couplings involved. The SM engineers breaking of the electroweak gauge group $SU(2)_L\times U(1)_Y\to U(1)_{\textrm{em}}$ via a Higgs field that develops a vacuum expectation value (vev) \cite{Higgs:1964pj}. The last missing link in the picture - the physical Higgs boson - that the SM predicts has now been discovered at the CMS \cite{Chatrchyan:2012xdj} \cite{CMS:yva} and ATLAS \cite{Aad:2012tfa} experiments. In spite of the various successes of the SM \cite{Aad:2013xqa} \cite{ATLAS:2013sla}, the nagging issue remains that there is no fundamental explanation for the scale of electroweak symmetry breaking (EWSB) within the SM. The Higgs field, while \emph{accommodating} EWSB, does not throw any light on the nature of EWSB itself. This is in stark contrast to the other scale in the SM, $\Lambda_{\textrm{QCD}}$, whose value can be inferred directly via the running of the strong coupling constant till the point where $g_{s}$ becomes large enough to bind quarks into color-singlet states. In addition, the SM does not have a candidate for dark matter nor can it explain neutrino masses. All such theoretical and phenomenological issues \cite{Heinemann:2019trx} have prompted theorists to try and construct models that extend the scope of the SM to address one or more of these questions -- these fall under the collective banner of ``Beyond the Standard Model" (BSM) theories.

BSM scenarios typically involving either enlarging the SM gauge group $SU(3)_C\times SU(2)_L\times U(1)_Y$ thereby invoking additional avenues of symmetry breaking, or enlarging the particle content with non-trivial charges under the SM gauge group, or both. A particularly attractive avenue along the latter lines is one that involves enlarging the scalar sector of the SM -- this can be done in a variety of ways  introducing additional Higgs fields that transform under the electroweak gauge group. The simplest of such models involves introducing an additional Higgs doublet, with both Higgs fields now participating in EWSB. This class of models, called the Two Higgs doublet models (2HDM) \cite{Branco:2011iw,Chiang:2013ixa,Bhattacharyya:2015nca,DeCurtis:2016tsm} can further be categorized depending on how the fermions couple to the two Higgs fields \cite{Haber:1978jt,Aoki:2009ha,Logan:2010ag}. This scalar spectrum of the so-called Type II 2HDM is identical to the Minimal Supersymmetric Standard Model (MSSM) \cite{Haber:1984rc}. In addition, one could also look for enlarged spectra with the Higgs field in representations other than the doublet under $SU(2)_L$. While such theories are typically constrained by a variety of theoretical and experimental factors \cite{Ross:1975fq}, there are many non-minimal representations that are phenomenologically interesting. Typically, since the higher representations include multiple scalar particles with non-trivial $T_3$ and $Y$ quantum numbers, one can typically expect, in addition to neutral scalars, singly or doubly charged Higgs bosons in such models. Examples include the Georgi-Machacek model \cite{Georgi:1985nv,Chanowitz:1985ug,Chiang:2015gva,Logan:2015xpa} which includes the SM Higgs doublet and in addtion, a real Higgs triplet with $Y=0$, supersymmetric models with extended Higgs sectors models \cite{Ellis:1988er,Drees:1988fc,Christensen:2013dra,Drees:1999sb,Kanemura:2014bqa}, Higgs triplets models \cite{Englert:2013zpa,Das:2016bir} that preserve $\rho=1$ at tree level \cite{Ross:1975fq}, and many more. \

In this paper, we undertake the collider study of the charged Higgs boson $\hpm$\cite{Akeroyd:2016ymd,Akeroyd:2016ssd, Maitra:2014qea} in a model independent fashion by categorizing BSM scenarios based on whether the charged Higgs is gaugophobic,chromophobic, or leptophobic - we lay out the essential details along with experimental inputs in Sec.~\ref{sec:overview} and identify the best case discovery modes for each. In Sec.~\ref{sec:analysis}, we present the collider study for some chosen benchmark points and then translate the discovery potential in the context of Type II 2HDM in Sec.~\ref{sec:model}. We present our conclusions in Sec.~\ref{sec:conclusions}. 

\section{Overview and Current Limits}
\label{sec:overview}
\subsection{Set-up and Strategy}
Enlarged scalar sectors in various BSM scenarios can in general have different gauge charges, and can also couple to fermions in the SM in different ways. While in principle many of these models also exhibit an enlarged gauge and/or matter spectrum, we restrict our attention to the simple case where the gauge group and fermionic content is purely SM-like. Even given this restriction, there are many possibilities for how the charged Higgs couples to the SM and any single study that hopes to encompass the myriad model building avenues that exist can only hope to do so by some form of broad classification of these models based on the nature of the charged Higgs couplings. In this spirit, we begin this study by analyzing three broad categories:
\begin{itemize}
\item Gaugophobic models: The charged Higgs has no couplings to the SM electroweak gauge bosons, particularly the $W^\pm$. 
\item Leptophobic models: The charged Higgs does not couple to the leptons in the SM.
\item Chromophobic models: Couplings of the charged Higgs to the colored particles in the SM are absent.
\end{itemize} 
While our aim here is not to present an overview of models that satisfy one or more of the above criteria, we note that realizations of the different cases can be easily understood. For instance, one could design an enlarged scalar spectrum with the Higgs multiplet containing the charged Higgs coupling only to leptons or quarks. Similarly, there are certain classes of deconstructed models  \cite{Chivukula:2009ck} in which the coupling of $\hpm$ to $\wpm$ and a scalar would be highly suppressed at tree level. Depending on the nature of its couplings, the charged Higgs will have rather different decay branching ratios (BR) and production mechanisms. Searches for the charged higgs have largely been restricted to its production via $gb\to H^{+}t$, or via top decay: $t\to H^{+}b$ if $m_{\hpm}<m_t$. In the former case, the predominant decay is to $tb$ while in the latter it could be $H^{+}\to \tau\nu$. While other channels like $A\wpm$ have been explored, to a large extent either the production or the decay have been one of the ``standard" cases. This is clearly untenable in a general search strategy if, for example, the charged Higgs is chromophobic and the $H^{+}tb$ vertex does not exist. Thus, at the outset, we would like to present the most viable channels in each model scenario and the rationale for the choices.

\begin{itemize}
\item \textbf{Gaugophobic}: The absence of any vertex of the form $\hpm A \wmp$\footnote{Here and in the rest of the paper, we will indicate generic heavy scalars by $A$ and $H$ (the typical symbols used in the 2HDM literature), and will reserve the symbol $h$ for the SM-125 GeV Higgs. In our study, we do not make use of angular correlations and hence will not distinguish between scalar and pseudoscalar decay modes explicitly.} means that the dominant decay modes are $\tau\nu$, $t\bar{b}$. Thus in this case we concentrate on the $gb\to H^{+}t$ production mode with $H^{+}\to t\bar{b}$. While $\tau\nu$ can certainly be considered, the purely hadronic mode aids in cleaner reconstruction (at the cost of higher backgrounds, of course).
\item \textbf{Chromophobic}: Since the $\hpm$ does not couple to colored particles, the production channel $pp\to \hpm t$ is absent and we need to look for the $\hpm$ as a decay product of a heavier particle like a heavy neutral scalar $H$. The possible $s$ channel mode $u\bar{d}\to H^{+}$ is suppressed by the small masses of the quarks, and hence would not be viable. Thus, in this case, we look at $pp\to H\to\wpm\hmp$ with $\hpm\to A\wmp$.
\item  \textbf{Leptophobic}: In this case, the production can be either $gb\to H^{+}t$ or $pp\to H\to\wpm\hmp$ with $\hpm\to A\wpm$ or $\hpm\to t\bar{b}$. Hence we will explore both the possibilities when analyzing this channel. 
We summarize all three cases in Table \ref{tab:classification}. 
\end{itemize}

\begin{table}[h!]
\centering
\begin{tabular}{ |c|c|c| }
\hline
\multicolumn{3}{ |c| }{Charged Higgs discovery modes} \\
\hline
Type & Production and Decay & Final state \\ \hline \hline
\multirow{1}{*}{Gaugophobic} &$pp \rightarrow H^{+} \bar{t}; H^{+} \rightarrow t b$ & $jjb\bar{b}\bar{b}\ell\nu$  \\ \cline{2-3}

  \hline\hline
\multirow{2}{*}{Chromophobic/Fermiophobic} 
 & $pp \rightarrow H \rightarrow W^{-} H^{+}; H^{+} \rightarrow W^{+} A$  &   
        $  jjb\bar{b}\ell\nu$ \\\cline{2-3}

 & $ pp \rightarrow H^{+} H^{-}; H^{\pm} \rightarrow W^{\pm} A$  & $4b + 2l+ 2\nu $ \\\cline{2-3}
        \hline\hline
        \multirow{1}{*}{Leptophobic} &$ pp \rightarrow H^{+} \bar{t}; H^{+} \rightarrow W^{+} A$ & $jjb\bar{b}\bar{b}\ell\nu$ \\ \cline{2-3} 
        & $ pp \rightarrow H \rightarrow W^{-} H^{+}; H^{+} \rightarrow t \bar{b}$ &   
        $ jjb\bar{b}\ell\nu$ \\\cline{2-3}
\hline
\end{tabular}
\caption{Possible production and decay modes of a charged Higgs boson in the three cases. In this paper, we will pick an optimal channel for each and detail the collider phenomenology for a few chosen benchmark points.}
\label{tab:classification}
\end{table}

It is seen that while the production and decay channels are quite different in the various scenarios, the final state for all of them contains multijets and $b$'s\footnote{In this paper, we will only consider single production channels of the charged Higgs. While pair production might be useful in certain models, here we would like to avoid the difficulties involved in reconstruction and the smaller cross-sections. }. However, the presence of $\ell+\met$ means that all SM backgrounds have at least one electroweak vertex\footnote{This is, of course, not an absolute necessity as there is also the possibility that there could be misidentified leptons etc. } - thus rendering the background small is less difficult compared to the scenario of a pure QCD background. On the other hand, we require high enough signal cross-sections that will withstand multiple $b$ tagging efficiencies and substantial $p_T$ cuts - we will see in the next section that in most cases, with stringent cuts the SM background can pretty much be nullified for many cases and thus these channels can be promising even if the signal cross-section is not too high.

While specific models that display the features of charged Higgs couplings displayed in Table \ref{tab:classification} can be interesting in their own right, as mentioned before we postpone such discussions and will present the phenomenology in a completely model-independent way as follows: we will pick the optimal channel for each class of models and do a signal vs background study for an optimal choice of cuts. The signal cross-section chosen here is arbitrary and the only goal here is to finalize a cut flow chart that suppresses the background without substantially affecting the signal cross-section. We will then use the number of background events left after imposing the cuts to back-calculate the signal cross-section necessary for a 5$\sigma$ discovery. In Sec.~\ref{sec:model}, we will do a \emph{model-dependent} analysis by translating our results in the parameter space of the Type II 2HDM. Specific models with an enlarged scalar sector have many constraints - both theoretical (perturbativity, unitarity) - and experimental ($\Delta\rho$, flavor constraints) that impose various relations between the masses of the new particles and the couplings. While a specific study should certainly cater to these constraints and filter out the parameter space in which to do the phenomenology, our goal here is to provide a sufficiently general analysis that is applicable to wide classes of models and hence in what follows we will treat the Higgs masses $m_{A},\,m_{\hpm}$ etc. in a typical multi-Higgs model as essentially independent parameters.

\subsection{Overview of the current experimental limits}

The ATLAS and CMS experiments have collected data independently from various phases of the collider run and have looked for a charged Higgs and thus far no conclusive evidence for the same has been found. Below, we collate the results of such findings and briefly discuss each result. One can categorize these search strategies into two cases depending upon the mass of charged Higgs: $m_{H^{\pm}}<m_{t}$ and $m_{H^{\pm}}$ $\geq$ $m_{t}$. The final state topology, and thus the search strategy, for these two cases is obviously different. Let us begin with the light $\hpm$ case.

\begin{itemize}
\item For $\hpm$ $\rightarrow$ c$\bar{s}$ channel, data has been collected during the different run phases with the integrated luminosity ranging from 4.7 fb$^{-1}$ to 19.7 fb$^{-1}$. From the combined analysis, the mass range 90 GeV - 160 GeV has been excluded \cite{Aad:2013hla} \cite{Khachatryan:2015uua}.
\item The CMS collaboration analyzed the data  they collected with $\sqrt{s}$ = 8 TeV and the integrated luminosity $\mathcal{L}$ = 19.7 fb$^{-1}$ for the decay channel $\hpm\rightarrow c\bar{b}$. No significant excess was found in the mass range 90 GeV - 150 GeV \cite{Sirunyan:2018dvm}. 
\item Multiple search analyses have been performed on the $\hpm\rightarrow \tau\nu_{\tau}$ channel during the different upgrades of the collider. The data which was collected for this study ranges in integrated luminosity from 2 fb$^{-1}$ to 35.9 fb$^{-1}$. The charged Higgs here is produced via top quark decay which in turn is produced in the $t\bar{t}$ production channel. The second top  (which did not decay to the $\hpm$)  would further decay to $\wpm b$ with the $\wpm$ further decaying to either leptonically or hadronically. The resultant mass exclusion for the case of light charged Higgs from both collaborations ranges from 80 GeV to 160 GeV \cite{Aad:2012tj} \cite{Aad:2012rjx} \cite{Chatrchyan:2012vca}.  
\end{itemize} 

For the case of the heavy charged Higgs, there are various production channels, i.e. associated production channel $p p \rightarrow H^{\pm}t$, VBF production process and $s$-channel production, that can each dominate depending on the mass and couplings of the $\hpm$.
\begin{itemize}
\item The charged Higgs which produced in the associated production process can further decay leptonically $\hpm\rightarrow\tau^{+}\nu_{\tau}$ \cite{Aad:2014kga} \cite{Khachatryan:2015qxa}, or to top-bottom pair $\hpm\rightarrow tb$ \cite{Aad:2015typ}. For the leptonic channel, data has been recorded for the integrated luminosity range 19.5 fb$^{-1}$ to 36.1 fb$^{-1}$ and masses in the range 180 GeV - 3 TeV have been excluded \cite{CMS:2016szv}. Further, the hadronically decaying $\hpm$ has been excluded in the range 200 GeV to 2 TeV \cite{Aaboud:2016dig} \cite{Aaboud:2018gjj} \cite{Aaboud:2018cwk}. 
\item The s-channel production process has been analyzed by the ATLAS collaboration \cite{Aaboud:2018ohp} for the integrated luminosity ranging from 20.3 fb$^{-1}$ to 36.1 fb$^{-1}$. In this case, the $\hpm$ further decays to $\wpm Z$. The two cases of the  electroweak gauge boson decaying semi-leptonically or fully leptonically have been analyzed and the charged Higgs mass range 400 GeV to 3 TeV has been excluded.
\item Vector Boson fusion can serve as another significant production channel for the case of heavy charged Higgs. Both ATLAS and CMS collaborations have collected data for integrated luminosities ranging from 15.2 fb$^{-1}$ to 20.3 fb$^{-1}$. $\hpm$ produced via the VBF process further decays to a $\wpm$ and a $Z$.  The CMS collaboration analyzed events in which both gauge bosons decayed leptonically \cite{Sirunyan:2017sbn}, whereas the ATLAS collaboration considered that $Z\rightarrow \ell^+\ell^-$ and $\wpm \rightarrow q q^{'}$ \cite{Aad:2015nfa}. The combined mass range which is excluded considering both the analyses ranges from 200 GeV to 2 TeV.  
\end{itemize}

While direct collider limits on the charged Higgs mass seem rather stringent, these limits should be interpreted within the context of \emph{specific} search strategies oftentimes assuming a 100\% BR to a desired channel. Thus, for instance, any search involving production or decay processes with a $tb$ would not apply to a chromophobic charged Higgs. Thus, in the next  sections we will proceed without unduly restraining the charged Higgs mass and analyze the collider phenomenology pertinent to the three broad categories discussed in the previous subsection.


\section{Collider Phenomenology}
\label{sec:analysis}

In this section we will analyse the $H^{\pm}$ search prospect for various discovery modes mentioned in Table \ref{tab:classification} at the 14 TeV LHC.  As mentioned in the previous section, this analysis will be done without recourse to a particular model in the sense that we will not be using any specific coupling or branching ratio (BR) values. However, we will make the following general assumptions about the kinds of models that our analysis applies to:
\begin{itemize}
\item The scalar spectrum of the model admits, in addition to a charged Higgs, additional neutral scalars (heavier than the SM Higgs). We will generically denote these states by $H$ and $A$, in keeping with the 2HDM/MSSM notation. In what follows, we will assume that $A$ is lighter than the $\hpm$ while the $H$ is heavier. 
\item We will not employ any specific $\mathcal{CP}$ properties of the $H$ and $A$, i.e., we will make no assumptions about whether they are scalars or pseudoscalars as we will not use any angular distribution analyses that will distinguish the two cases.\footnote{However we point out for clarity that we do not include any couplings that is disallowed by $\mathcal{CP}$ symmetry.}
\item For a specific case, say Chromophobic, we will assume that all couplings of the $\hpm$ other than those to colored particles allowed by the SM symmetries are indeed present. This will simplify the analysis as we do not need to make too many model-specific assumptions.
\end{itemize}

To proceed, we choose three benchmark points $m_{H^{\pm}}$ = 300 GeV, 500 GeV, 700 GeV to perform the analysis. The generic production modes of the $\hpm$ that we will consider are through the decay of the $H$ and the associated production with a top-quark. In order to be left with sufficient number of signal cross-section after the cuts, we have chosen low/moderate values of $m_{\hpm}$ particularly for the case where it is the decay product of a heavy $H$. In addition, we fix the masses of the $H$ and $A$ to be 800 GeV and 150 GeV respectively. We performed the data simulation using the {\sc MadGraph}5\_{aMC@NLO} \cite{Alwall:2014hca}  event generator. The SM backgrounds which are used for this study are generated via the in-built SM model file in the  {\sc MadGraph} repository. To generate the signal distributions, we built a BSM model file in {\sc FeynRules} \cite{Christensen:2009jx} \cite{Alloul:2013bka} with an enlarged scalar sector as detailed in the assumptions above. The parton level simulation from  {\sc MadGraph} were then passed on to the {\sc Pythia}~6 \cite{Sjostrand:2006za} program for showering and hadronization. Detector level simulations of the resulting events were performed with {\sc Delphes}~3 \cite{deFavereau:2013fsa} and the ensuing objects were reconstructed employing the {\sc MadAnalysis}~5 \cite{Conte:2012fm} \cite{Conte:2014zja} framework which was also used to perform our cut-based analysis that is detailed in the forthcoming sections.

As explained in the preceding sections, we separate the signal into different classes based on the peculiarities of the charged Higgs coupling. While one could fine tune the phenomenological analysis in each case to cater to its own peculiarities, it is more profitable to exploit the commonalities in the different scenarios so the search strategy is not greatly different. To do so, we first note that the signals presented in Table~\ref{tab:classification} can be classified into two classes based upon the exclusive partonic final states available through the decay cascade of the heavier particles produced at LHC:  (2$j$ + 2$b$ + $\ell\nu$) and (2$j$ + 3$b$ + $\ell\nu$). 

\begin{table}[h!]
\centering
\begin{tabular}{ |c|c|c|c|m{5 cm}| }
\hline
Signal & Chromophobic & Gaugophobic & Leptophobic & Remarks \\ \hline
2$j$ + 2$b$ + $\ell\nu$ & \cmark & \xmark & \cmark &Identical production mode. $\hpm\to\wpm A$ in Chromophobic while $\hpm\to tb$ in Leptophobic. \\ \hline
2$j$ + 3$b$ + $\ell\nu$ & \xmark & \cmark & \cmark &Identical production mode. $\hpm\to\wpm A$ in Leptophobic while $\hpm\to tb$ in Gaugophobic. \\ \hline
\end{tabular}
\caption{Classification of signals based on final state topology. The final spectrum has similar kinematic properties in the two cases.}
\label{tab:final_state}
\end{table}

In addition to having identical final states, the particles themselves have similar kinematic properties in the two cases because of a common production mode with differences being introduced because the decay of the $\hpm$. We see that the leptophobic charged Higgs can be looked for in both the channels owing to its unsuppressed $tb$ couplings and gauge interactions. Also, in keeping with Table~\ref{tab:classification}, one could also look for pair production of the charged Higgs in the chromophobic scenario leading to a different final state from those tabulated above, but we do not pursue it here.

Given the multijet final state, the major experimental search challenges come from dominant SM process like $t\bar{t}$+jets and $WZ$+jets\footnote{Since the signal events have only one source of missing energy, we neglect $ZZ$+ jets background. However, we have checked that this background gives negligible contribution once we impose the set of cuts devised.}. The presence of $l\nu$ in the signal final state helps in suppressing a large number of pure QCD background events, particularly for signals with appreciable lepton $p_T$. To begin, we employ the following set of basic identification cuts at the time of simulation to help eliminate any soft jets and leptons: 
\begin{equation}
p_T^j > 20~{\rm GeV}, \qquad p_T^\ell > 10~{\rm GeV}, \qquad
  |\eta^j|\leq5\qquad\text{and}\qquad  |\eta^\ell| \leq 2.5 \ .
\label{eq:basic1}
\end{equation} 
We have chosen a wider window for the pseudorapidity for jets as compared to the leptons to ensure that we do not lose many signal events. Further, we demand that all pairs of particles are optimally separated:
\begin{equation}
\Delta R_{jj}=\Delta R_{bb}=\Delta R_{jl}=\Delta R_{bj}=0.4.
\label{eq:basic2}
\end{equation}
With this basic framework now in place, we now turn our attention to optimizing the discovery process of the charged Higgs by designing kinematic cuts for the two different final states.

\begin{figure}[h!]
\includegraphics[scale=0.4]{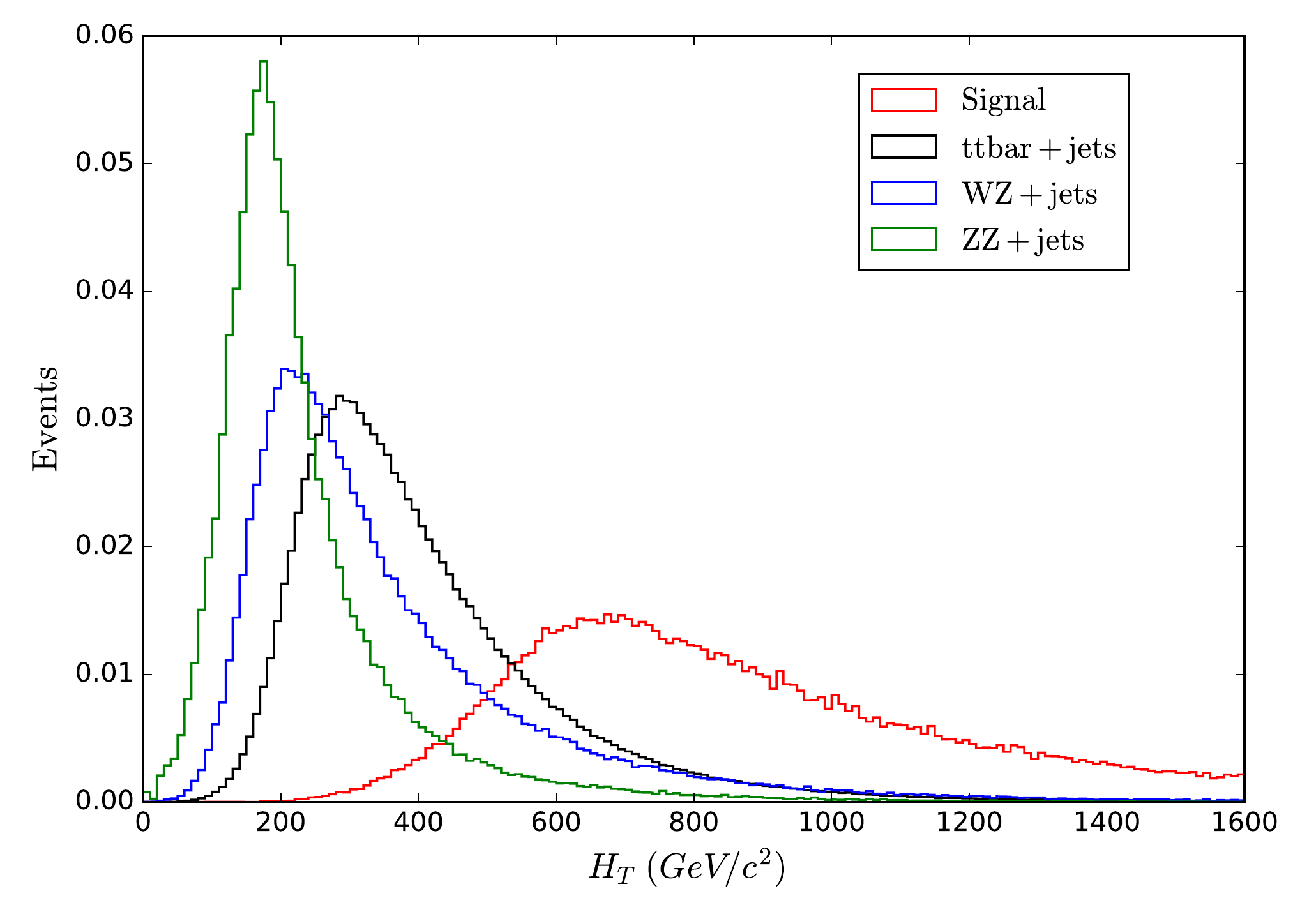}
\caption{The $\ensuremath{{\not\mathrel{H}}_T}$ distribution for both signal and background for chromophobic case for the case of a 500 GeV charged Higgs. The plot shows that SM background events cluster mostly below 400 GeV in contrast to the signal.}
\label{fig:chromo_THT}
\end{figure}

\begin{figure}[h!]
\includegraphics[scale = 0.4]{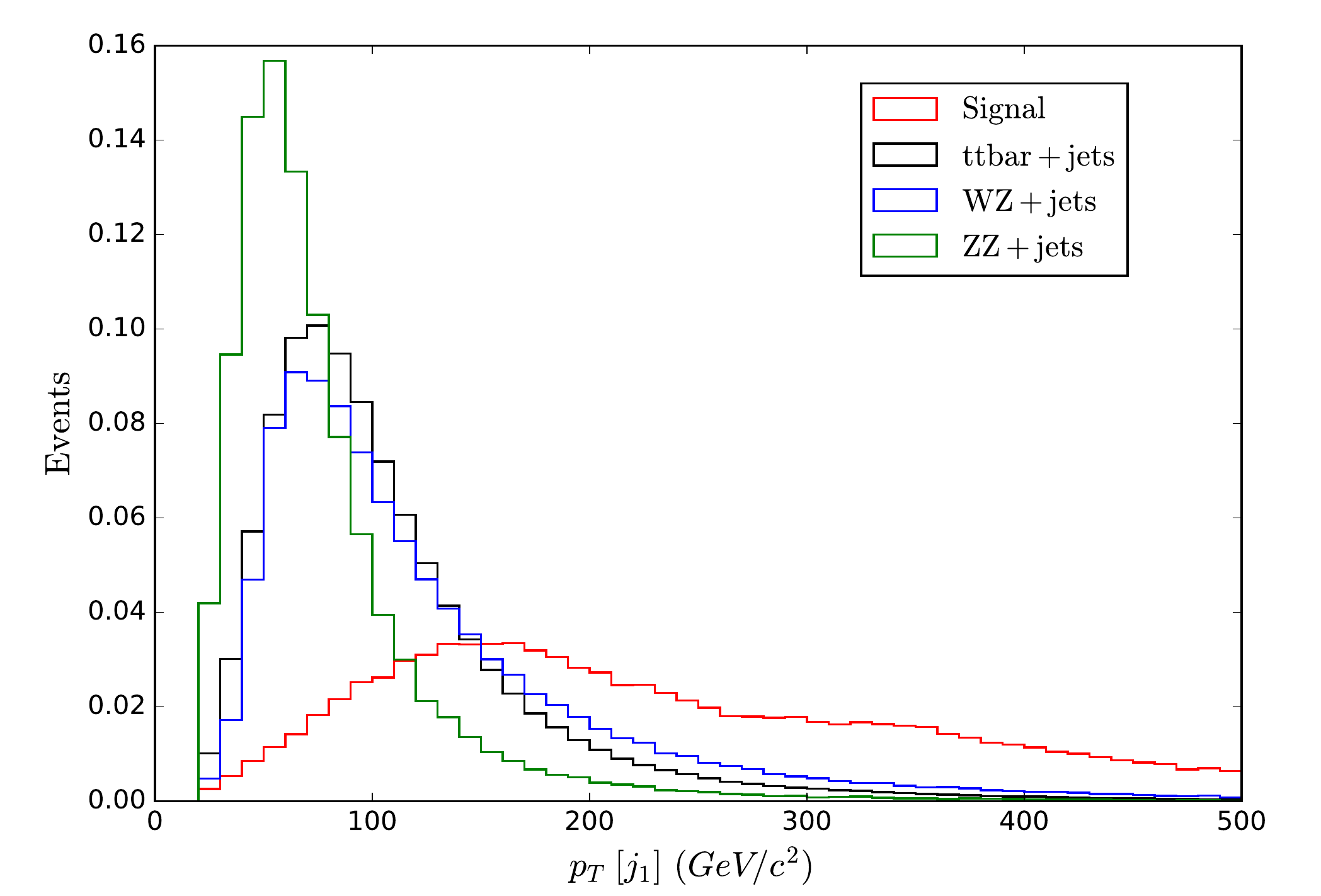}\hspace{0.1in}
\includegraphics[scale = 0.4]{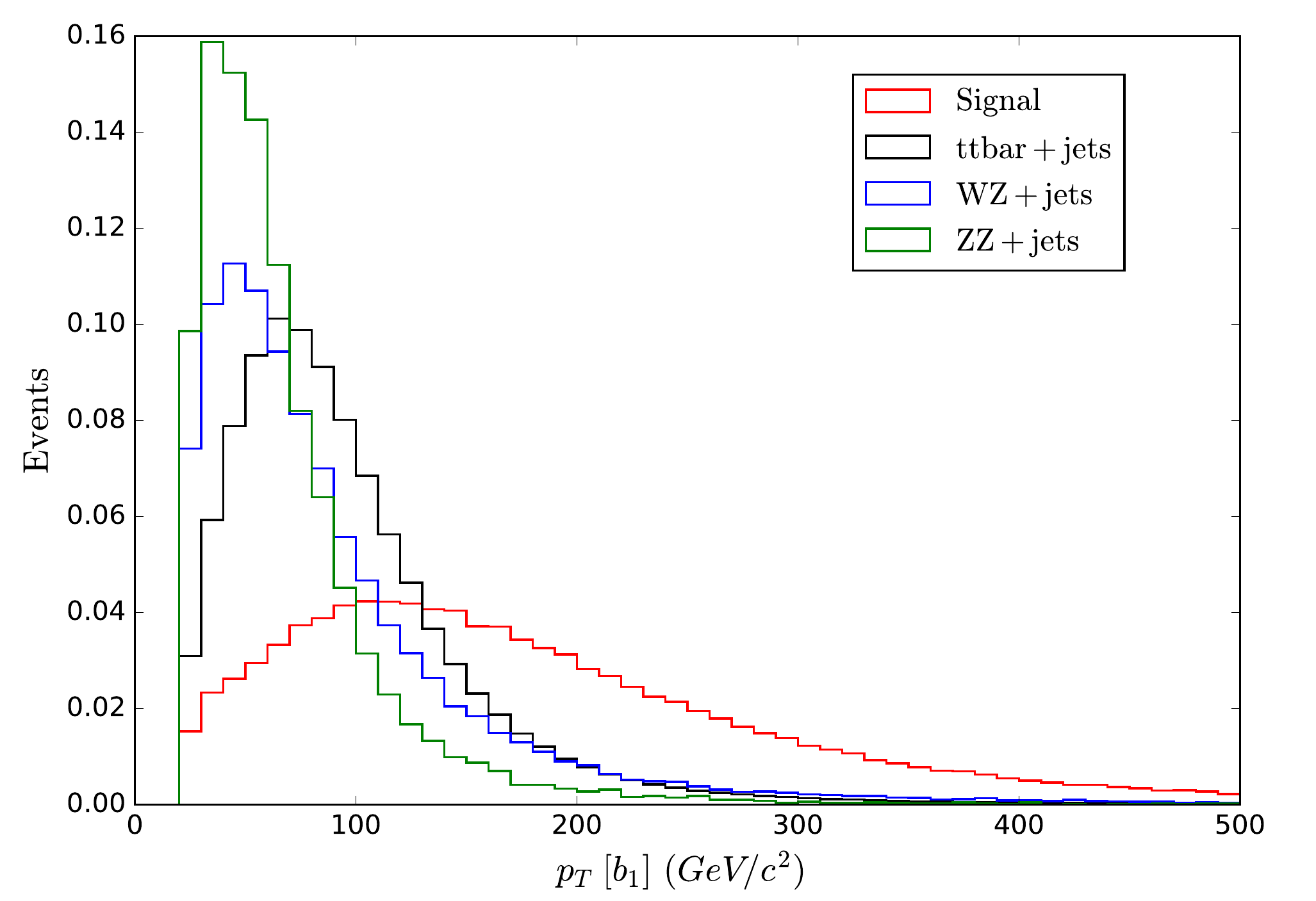}
\caption{The $p_T$ distribution for the leading four flavor jet  (left)and the leading $b$ jet (right). The benchmark point of the distribution is $m_{\hpm} = 500$ GeV.}
\label{fig:pT_chromo}
\end{figure}

We begin with the 2$j$ + 2$b$ + $\ell\nu$ channel -- as can be seen in Table~\ref{tab:final_state}, this applies to both the chromophobic and leptophobic channels. We employ the self-evident set of identification cuts: $N(j) \geq 2, N(b) = 2$  and $N(\ell) = 1$. As can be seen from Table \ref{tab:2b2j}, this will reduce more than 90\% of the background but as collateral damage, we do also lose a large number of signal events. We note at this stage that the signal cross-section numbers given in Table~\ref{tab:2b2j} is fiducial in nature -- the purpose of this table is to simply illustrate the efficacy of the cut flow, i.e., systematically eliminate the background without unduly reducing the signal. We reserve all model-specific implications to Section \ref{sec:model}. The first kinematic quantity which we use to eliminate the background is the total transverse hadronic energy $H_{T}$. In the signal all the hadronic particles are produced via the decay of heavily boosted mother particles unlike its SM counterpart. As a result, in Fig.~\ref{fig:chromo_THT}, one can notice a wider spread in the signal events (shown for $m_{\hpm}=500$ GeV) -- this prompts us to choose $H_{T} \geq $ 400 GeV to eliminate the SM background\footnote{Since the distributions are rather similar in both the chromophobic and leptophobic scenarios, we present only the plots for chromophobic case for illustration purposes.}. While the $p_T$ distribution of the leading jets also show a somewhat similar behavior qualitatively (see Fig.~\ref{fig:pT_chromo}), i.e., the signal has a longer tail while the SM is peaked at lower $p_T$ values, imposing hard $p_T$ cuts would run the risk of losing an increased number of signal events in this case. Thus we choose the conservative cuts $p_{T}(j_{1}) \geq 75$ GeV and $p_{T}(b_{1}) \geq 75$ GeV to achieve an enhanced $\frac{S}{\sqrt{B}}$ ratio, with the $p_T$'s of the subleading jets constrained only by the initial selection cuts. 

\begin{figure}[h!]
\includegraphics[scale=0.4]{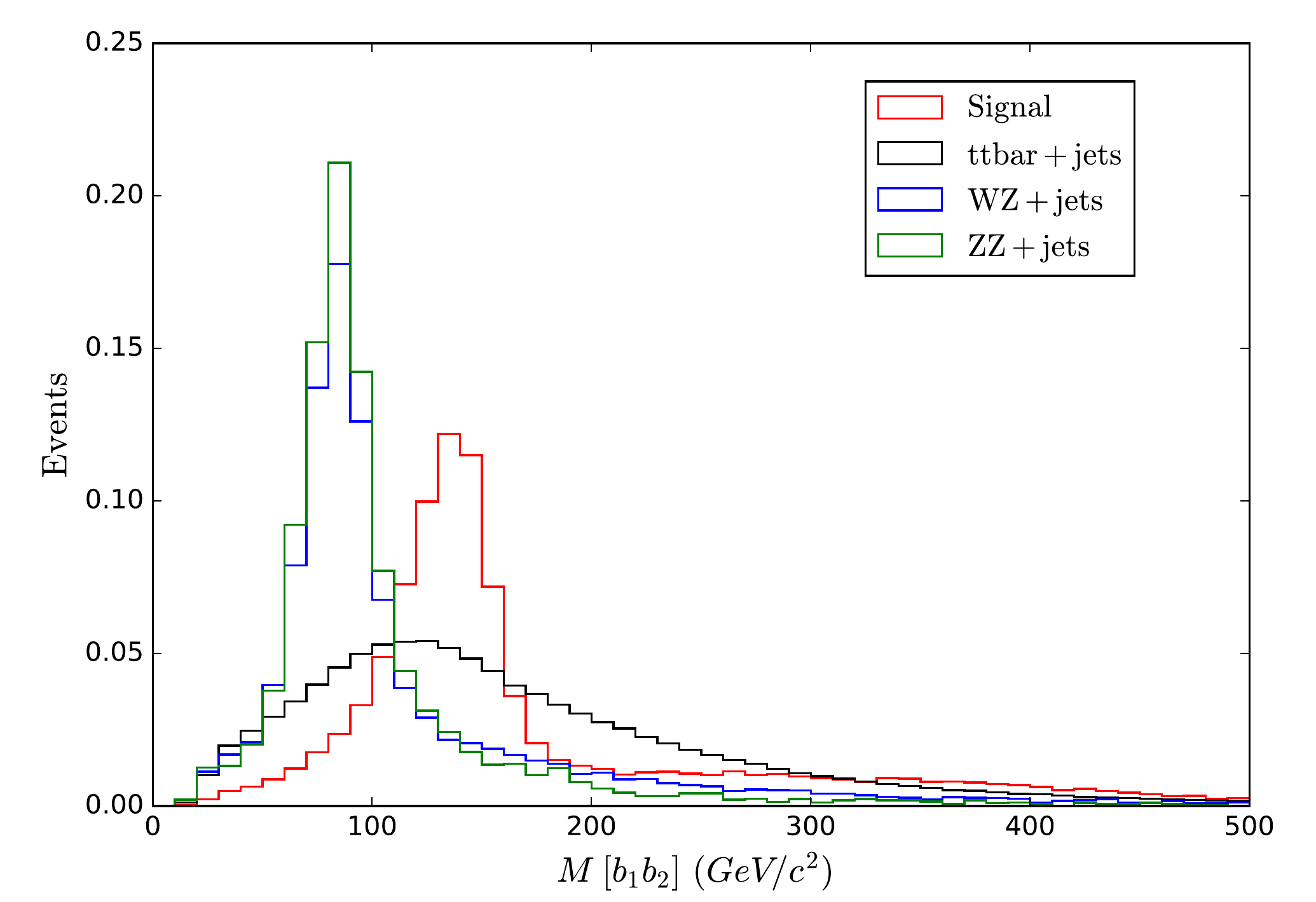}
\hspace{0.1in}
\includegraphics[scale=0.4]{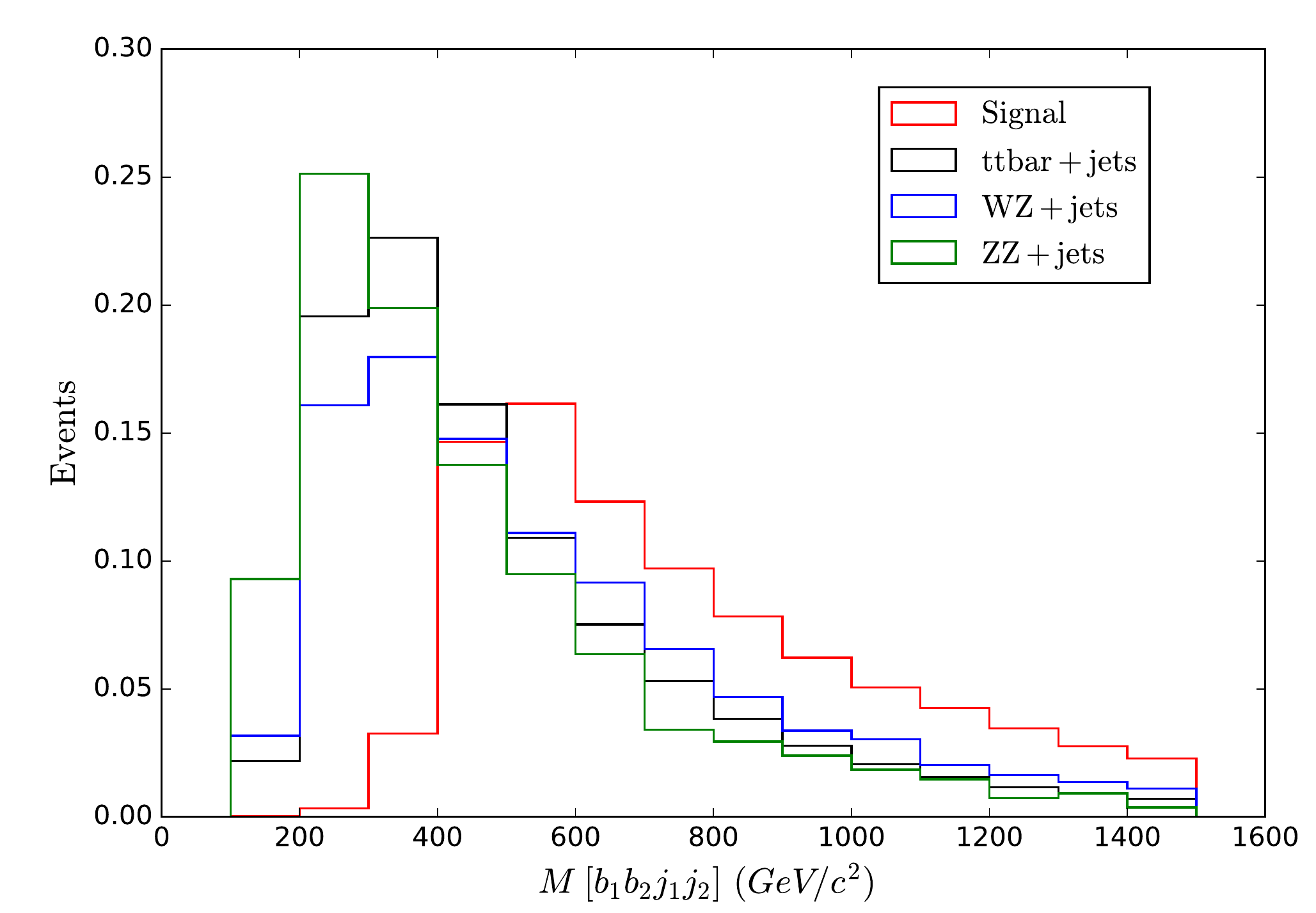}
\caption{Invariant mass distributions of $m_{bb}$ (left) and $m_{bbjj}$ (right) for the chromophobic case for a 500 GeV $\hpm$. It is seen that the $m_{bb}$ distribution for the signal clearly peaks around $m_h=125$ GeV, while the $m_{bbjj}$ peaks around $m_{\hpm}$ motivating the cuts given in Table~\ref{tab:2b2j}.}
\label{fig:invmass_chromo}
\end{figure}

Finally having thus isolated the signal, we turn to the final step of employing suitable invariant mass cuts. Notice that  the chromophobic signal involves a pair of $b$ jets originating from the 150 GeV Higgs, and the $2b+2j$ combination that should reconstruct the charged Higgs. In the leptophobic case, only the latter condition is true as the two $b$ jets come from $\hpm$ and top decay. In Fig.~\ref{fig:invmass_chromo}, we display both these for the chromophobic signal and background. It is seen that the SM background to $b\bar{b}$ from $VV+$jets understandably predominantly arises from the decay of the $Z$ while the $t\bar{t}$+jets leads to a smoother distribution as the $b$'s in this case result from the hadronic decays of a boosted $W$. Based on these observations, we choose the cuts $120\,\textrm{GeV}  \leq m_{bb} \leq 180\,\textrm{GeV} $ and  $(m_{\hpm} -100)\,\textrm{GeV} \leq m_{bbjj} \leq (m_{\hpm} +100)\,\textrm{GeV}$. We note that in the former case we have deliberately chosen a rather asymmetrical cut both to eliminate the background and not lose too much signal in the process based on the distributions in Fig.~\ref{fig:invmass_chromo}. For the leptophobic case, only the $m_{bbjj}$ cut is applied. This is reflected in Table~\ref{tab:2b2j} as well: in the last column, the numbers in bracket correspond to the background events left after the $m_{bbjj}$ cut when the previous $m_{bb}$ cut is \emph{not} applied.

We present the cutflow chart for the 2$j$ + 2$b$ + $\ell\nu$ channel for both the chromophobic and leptophobic scenarios choosing $m_{\hpm}=500$ GeV in Table~\ref{tab:2b2j} imposing the cuts discussed in the preceding paragraphs - we see that the progressive kinematic cuts have done a good job in systematically suppressing the SM background. In Table~ \ref{tab:3b2j}, we present the corresponding numbers for the 2$j$ + 3$b$ + $\ell\nu$ channel for the gaugophobic and leptophobic cases -- one can see a similar trend of suppression of the SM in the case as well.

\begin{table}[]
\centering
\begin{tabular}{|l|l|l|l|l||l|l|}
\hline
\multirow{2}{*}{Cut} & \multicolumn{2}{c|}{Background} & \multicolumn{4}{c|}{Signal}            \\ \cline{2-7} 
                     									& $t\bar{t}$+jets            & $WZ$+jets             & Chromophobic &$\frac{S}{\sqrt{B}}$   & Leptophobic & $\frac{S}{\sqrt{B}}$  \\ \hline
 Initial      															    &  2000000              &    500000            & 100000             &\,\, \,-    &   100000          & \,\, \,-    \\ \hline
 $N_j$ $\ge$ 2          													    &    1894836            & 475313               & 97053             &63.04     & 97005            &  65.85   \\ \hline
 $N_l$ = 1               													    &     539543            &274654                &69461              &76.97     &  69590           & 77.12    \\ \hline
 $N_b$ = 2       													             &     163773           &11002                & 24310             & 58.14    & 25097            &60.03     \\ \hline
 $H_T \ge 400$ GeV    												     &    83444             &6124                & 24038              & 80.32    & 24798            & 82.86    \\ \hline
 $p_T(j_1) \ge 75$ GeV                                                                                                     &     70543            &5531                &  22468            &81.46     & 21769            & 78.92    \\ \hline
 $p_T(b_1) \ge 75$ GeV                                                                                                    &    56828            & 4267               &20998              & 84.95    & 20740             &83.91     \\ \hline
 $120$ GeV$ \leq m_{bb} \leq$ 180 GeV                                                                      &    11716            & 477               & 11716             & 93.24    & NA            &   NA  \\ \hline
 $(m_{\hpm} -100)\,\textrm{GeV} \leq m_{bbjj} \leq (m_{\hpm} +100)\,\textrm{GeV}$         &      5893 (20280)         &  186 (1227)             & 4333             & 55.56    &  6642           &45.29     \\ \hline               
\end{tabular}
\caption{Cut flow chart for the 2$j$ + 2$b$ + $\ell\nu$ channel with the signal corresponding to a 500 GeV $\hpm$ in both the chromophobic and leptophobic cases. In the last row, the numbers in bracket correspond to the background when the previous cut on $m_{bb}$ is not applied -- this is relevant to the leptophobic case.}
\label{tab:2b2j}
\end{table}

\begin{table}[h!]
\centering
\begin{tabular}{|l|l|l|l|l||l|l|}
\hline
\multirow{2}{*}{Cut} & \multicolumn{2}{c|}{Background} & \multicolumn{4}{c|}{Signal}            \\ \cline{2-7} 
                     									& $t\bar{t}$+jets            & $WZ$+jets             & Gaugophobic &$\frac{S}{\sqrt{B}}$   & Leptophobic & $\frac{S}{\sqrt{B}}$  \\ \hline
 Initial      															    &  2000000              &    500000            & 100000             &\,\, \,-    &   100000          & \,\, \,-    \\ \hline
 $N_j$ $\ge$ 2          													    &    1894836            & 475313               & 97017             &63.02     & 97232            &  63.15   \\ \hline
 $N_l$ = 1               													    &     539543            &274654                &50763              &56.25     &  51620           & 57.2    \\ \hline
 $N_b$ = 3       													             &     21564           &821                & 13013           & 86.97    & 12987            &86.8     \\ \hline
 $H_T \ge 400$ GeV    												     &    14800             &613                & 12691              & 102.22    & 12702            & 102.3    \\ \hline
 $p_T(j_1) \ge 75$ GeV                                                                                                     &     11204            &487                &  10533            &97.41     & 11634            & 107.6    \\ \hline
 $p_T(b_1) \ge 75$ GeV                                                                                                    &    9723            & 383               &10175              & 101.21    & 11153             &110.94     \\ \hline
 $(m_{\hpm} -100)\,\textrm{GeV} \leq m_{bbjj} \leq (m_{\hpm} +100)\,\textrm{GeV}$         &    4734         &  94             & 4734             & 77.29    &  4458           &72.79     \\ \hline               
\end{tabular}
\caption{Cut flow chart for the 2$j$ + 3$b$ + $\ell\nu$ channel with the signal corresponding to a 500 GeV $\hpm$ in both the gaugophobic and leptophobic cases.}
\label{tab:3b2j}
\end{table}

Having thus performed a largely model-independent analysis, we now turn to the issue of how large a cross-section a particular model should have in order for the charged Higgs to be discoverable using the methods outlined above. It is simple enough to take the background events in each case, and estimate the actual number of signal events necessary to obtain a 5$\sigma$ discovery -- these numbers are presented for the various scenarios (and for different benchmark points) in Table~\ref{tab:cross-section}. We now turn to the question of realizability of these numbers in the context of a specific model.

\begin{table}[h!]
\centering
\begin{tabular}{|c|c|c|c|c|c|}
\hline
\multirow{2}{*}{Production Channel} & \multirow{2}{*}{Benchmark Points} & \multicolumn{2}{c|}{2$\sigma$ Significance} & \multicolumn{2}{c|}{5$\sigma$ Significance}            \\ \cline{3-6} 
                      									&              & $\mathcal{L}$ = 500 fb$^{-1}$ & $\mathcal{L}$ = 1000 fb$^{-1}$   & $\mathcal{L}$ = 500 fb$^{-1}$& $\mathcal{L}$ = 1000 fb$^{-1}$  \\ \hline
               & 300 GeV & 2.997 & 2.073 & 7.526 & 5.31 \\
Chromophobic &       &          &           &         &        \\
               		& 500 GeV & 2.75 & 1.939 & 6.899 & 4.866 \\
		\hline
               & 300 GeV & 2.707 & 1.912 & 6.792 & 4.767 \\
Gaugophobic &       &          &           &         &        \\
               		& 500 GeV & 2.447 & 1.686 & 6.151 & 4.339 \\
		\hline
               & 300 GeV & 2.707 & 1.912 & 6.792 & 4.767 \\
Leptophobic ($2j+3b+\ell\nu$) &       &          &           &         &        \\
               		& 500 GeV & 2.447 & 1.686 & 6.151 & 4.339 \\
		\hline
               & 300 GeV & 5.376 & 3.8 & 13.465 & 9.516 \\
Leptophobic ($2j+2b+\ell\nu$) &       &          &           &         &        \\
               		& 500 GeV & 5.142 & 3.629 & 12.995 & 9.11 \\
		\hline													
  \end{tabular}
\caption{The cross-sections required for the 5$\sigma$ and 2$\sigma$ exclusion of the $m_{H^{\pm}}$ for the different signal scenarios detailed in Tables~\ref{tab:2b2j} and \ref{tab:3b2j} for different values of integrated luminosity.}
\label{tab:cross-section}
\end{table}


\section{Model Implications}
\label{sec:model}
\subsection{Cross-sections and Couplings}
In the previous sections we have detailed the collider phenomenology of the charged Higgs looking at various production and decay channels (see Table~\ref{tab:classification} for a quick summary). Combining the various classes of signals based on the final states, we have chosen a set of optimal cuts which help to reduce the corresponding SM background. The number of background events remaining after this set of cuts was then used to back-calculate the signal strength necessary for a 5$\sigma$ discovery - these details are presented in the Table~\ref{tab:cross-section}. To truly ascertain the efficacy of the approach, one needs to analyze the feasibility to realize the signal cross-section in a particular model with an enlarged scalar spectrum. While one should, strictly speaking, use models in which the $H^{\pm}$ is leptophobic, chromophobic, or gaugophobic and compare with the corresponding cross-section numbers, our goal here is not to do an overview of models. Thus, we choose a simpler strategy of choosing a particular model - the Type II Two Higgs-double Model (2HDM) - and turning off the couplings to leptons, colored particles, or gauge bosons to do the comparison in the three cases of interest. The relevant couplings in this model are displayed in Table~\ref{tab:couplings}. While we reiterate that the 2HDM does not fall in any of these classes, this analysis should give a sense of the numbers involved, and the efficacy of the cuts in each case.

\begin{table}[h!]
\centering
\begin{tabular}{|c|c|}
\hline
Vertex & Coupling\\
\hline
$g_{H^{\pm}W^{\mp}h}$ & $-\frac{ig}{2}\cos(\beta - \alpha)$\\
\hline
$g_{H^{\pm}W^{\mp}A}$ & $\frac{g}{2}$ \\
\hline
$g_{H^{\pm}qq^{'}}$ & $\frac{ig}{2\sqrt{2}m_{W}}[(m_{q^{'}}\tan\beta + m_{q}\cot\beta) - (m_{q^{'}}\tan\beta - m_{q}\cot\beta)\gamma_{5}]$\\
\hline
$g_{H^{\pm}\tau\nu}$ & $\frac{ig}{2\sqrt{2}m_{W}}[m_{\tau}\cot\beta(1 - \gamma_{5})]$\\
\hline
\end{tabular}
\caption{The relevant couplings between the charged Higgs and the quarks, leptons and bosons in the Type II 2HDM.}
\label{tab:couplings}
\end{table}

The regions of parameter space that admit a 5$\sigma$ discovery using the methods outlined in the previous section will obviously depend on the production cross-section of the $\hpm$ and its branching ratios to the relevant final states. While the branching ratios for each case need to be calculated separately, for the purposes of efficient organization of the results, it is useful to note that the charged Higgs in both classes of signals is produced via either associated production (gaugophobic and leptophobic cases) or as the decay product of a heavier scalar $H$ (chromophobic and leptophobic cases). We will briefly describe the two channels before moving on to the study of the parameter spaces.

\begin{figure}[ht!]
\begin{center}
\includegraphics[scale=0.6]{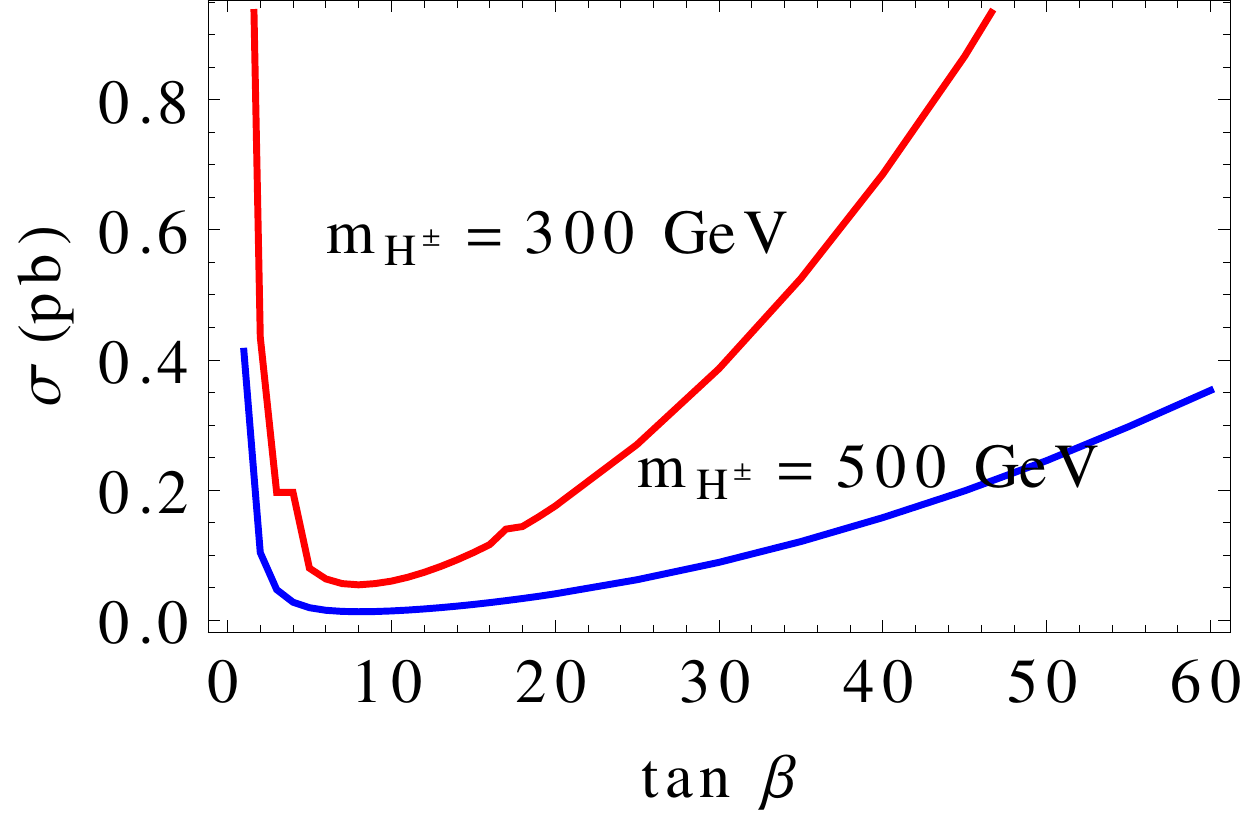}
\caption{The plot shows the cross-section versus $\tan\beta$ for the case of associated production of charged Higgs $\sigma$($gb \rightarrow H^{\pm}t$) for $m_{H^{\pm}}$ = 300 GeV and 500 GeV. The cross-section is enhanced for small and large values of $\tan\beta$ and flattens out in the region around $\tan\beta\approx$ 7. }
\label{fig:asso}
\end{center}
\end{figure}

Fig[\ref{fig:asso}] shows the cross-section as a function of  $\tan\beta$ for the associated production process, $\sigma(gb \rightarrow H^{\pm}t)$ at the 14 TeV LHC. The required cross-section values for different charged Higgs mass, $m_{H^{\pm}}$ = 300 GeV and $m_{H^{\pm}}$ = 500 GeV is collected from the report published by LHC working group [\cite{Dittmaier:2009np}]. Referring to Table \ref{tab:couplings}, it is seen that the cross-section determined by $g_{H^{\pm}tb}$ will be enhanced for both small and large values of $\tan\beta$ because of the presence of both $\tan\beta$ and $\cot\beta$ terms -- this is borne out by the plot, wherein one can see the enhancement in cross-section in the regions $\tan\beta<7$ and $\tan\beta>20$.  The region $\tan\beta$ $\approx$ 7 affords no such enhancement and is typically the region that is difficult to probe in charged Higgs searches in conventional channels. Further, for more massive $\hpm$, the enhancement in large $\tan\beta$ region is not as pronounced -- thus, in this case, one needs to choose the decay channel of $H^{\pm}$ pragmatically such that the signal has a high value of $\sigma$$\times$BR -- we will revisit this issue in the subsequent sections.

The second channel of interest in our study is the production of the $\hpm$ from the decay of a heavy neutral Higgs $H$. In order to calculate the cross-section $\sigma(gg \rightarrow H)$, one can always use the corresponding SM production cross-section by a suitably rescaled loop factor \cite{Djouadi:2005gi,Djouadi:2005gj}:
\begin{equation}
\sigma_{2HDM}(gg \rightarrow H) = \frac{\sigma_{SM}|\frac{\sin\alpha}{\sin\beta}F^{h}_{1/2}(\tau_{t}) + \frac{\cos\alpha}{\cos\beta}F^{h}_{1/2}(\tau_{b})|}{|{F^{h}_{1/2}(\tau_{t})} + F^{h}_{1/2}(\tau_{b})|},
\label{eqn:scaling}
\end{equation}
where $\tau_{f}$ = $\frac{4m_{f}^{2}}{m_{H}}$ (with $f = t, b$) and loop factor is $F^{h}_{1/2}$ = -2$\tau[1 + (1 - \tau)f(\tau)]$ and
\begin{equation}
	f(\tau) = \left\{ \begin{array}{lc} 
	\left[ \sin^{-1} (1/\sqrt{\tau}) \right]^2 \ \ & \tau \geq 1 \\   
	-\frac{1}{4} \left[ \ln\frac{1 + \sqrt{1 - \tau}}{1 - \sqrt{1 - \tau}} - i \pi \right]^2 \ \ & \tau < 1.
	\end{array} \right.
\end{equation}

\begin{figure}[ht!]
\begin{center}
\includegraphics[scale=0.5]{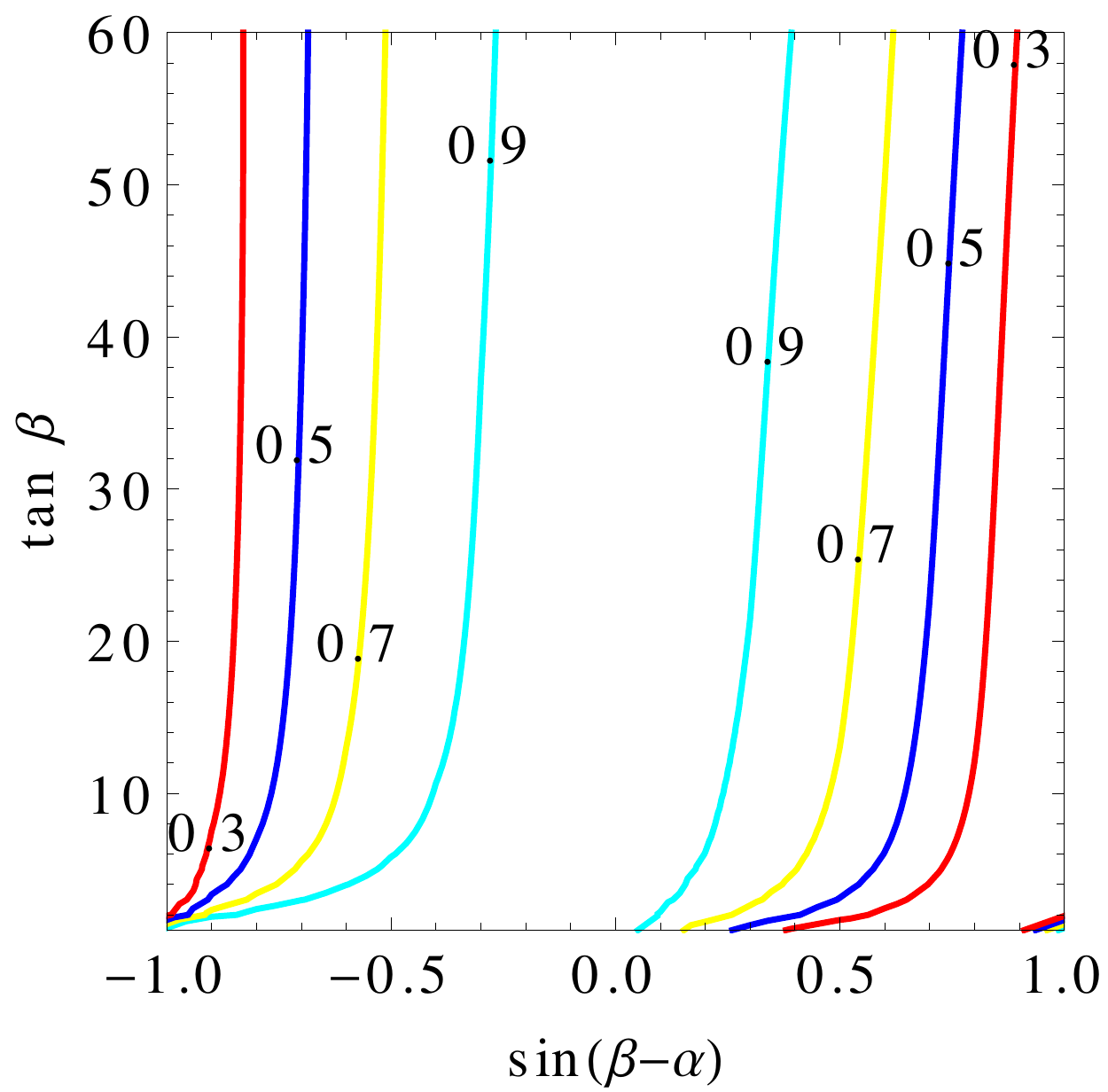}
\caption{{The plot shows the dependence of production cross-section of $\sigma(gg \rightarrow H^{0})$ with the free parameter ($\sin(\beta - \alpha), \tan\beta$). The neutral Higgs boson mass $H^{0}$ has set to $m_{H^{0}}$ = 800 GeV.}}
\label{fig:Gluon}
\end{center}
\end{figure}

In Fig.~\ref{fig:Gluon} we show the dependence of the gluon fusion production cross-section with the underline parameter plane $\sin(\beta - \alpha)$ versus $\tan\beta$ fixing $m_{H}$ = 800 GeV. The plot shows the contours of $\sigma_{H}/\sigma_{SM}$, where $\sigma_{SM}$ is the cross-section of the corresponding SM Higgs.  The cross-section is maximal near the $\sin(\beta - \alpha) \approx 0$ regions and does not show appreciable dependence on $\tan\beta$ unlike the associated production channel.  The plot is not completely symmetrical about $\sin(\beta - \alpha)=0$ -- this asymmetry arises due to the complicated loop factors in Eqn.~\ref{eqn:scaling}.  
\subsection{Discovery and Exclusion Regions}
With the basic structure now in place, we turn to the final question of analyzing the parameter space in the Type II 2HDM (with appropriate coupling modifications as discussed earlier) that would permit a 5$\sigma$ discovery or a 2$\sigma$ exclusion. We will do this for the three scenarios separately and comment on the results. We point out at the outset that there are many constraints on this model on both the theoretical (vacuum stability, perturbativity etc.) and experimental (observation of the 125 GeV Higgs, $\Delta\rho$, $b\to s\gamma$ etc.) fronts, and these together constrain the available parameter space of the model. A complete analysis of all such constraints is beyond the scope of this paper (see for example \cite{Coleppa:2013dya}), and thus we present the discovery and exclusion regions on the entire parameter space of Type II 2HDM. However, one should note that some of this parameter space might already be ruled out owing to the aforementioned considerations. However, our aim here is to try and understand the maximal available discovery regions for the particular collider analysis detailed in the previous section.

\subsubsection{Gaugophobic Models}
\begin{figure}[ht!]
\begin{center}
\includegraphics[scale=0.65]{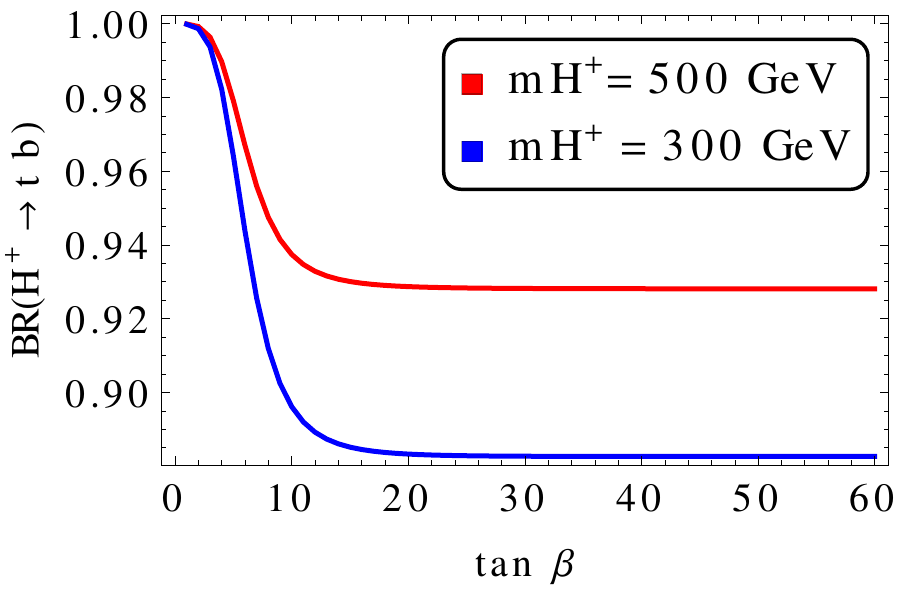}
\caption{The plot of the branching ratio of the $H^{\pm}$ in the $t\bar{b}$ channel for $m_{H^{\pm}}$ = 500 GeV (red) and $m_{H^{\pm}}$ = 300 GeV (blue). The BR is maximal for all values of $\tan\beta$ owing to the absence of the other channels.}
\label{fig:gaugo_tb}
\end{center}
\end{figure}

Here, the $H^{\pm}$ is produced via associated production and decays predominantly to $t \bar{b}$ -- thus, the $g_{H^{\pm}tb}$ coupling plays a crucial role. In Fig[\ref{fig:gaugo_tb}] we show the branching ratio (BR) in the $tb$ channel as a function of           $\tan\beta$ for two different charged Higgs masses. It is clear that the absence of the $AW^\pm$ and $hW^\pm$ channels has significantly enhanced this BR and it is more than $\approx 90\%$ in the entire parameter space with the only competing channel being $\tau\nu$.

\begin{figure}[ht!]
\begin{center}
\includegraphics[scale=0.4]{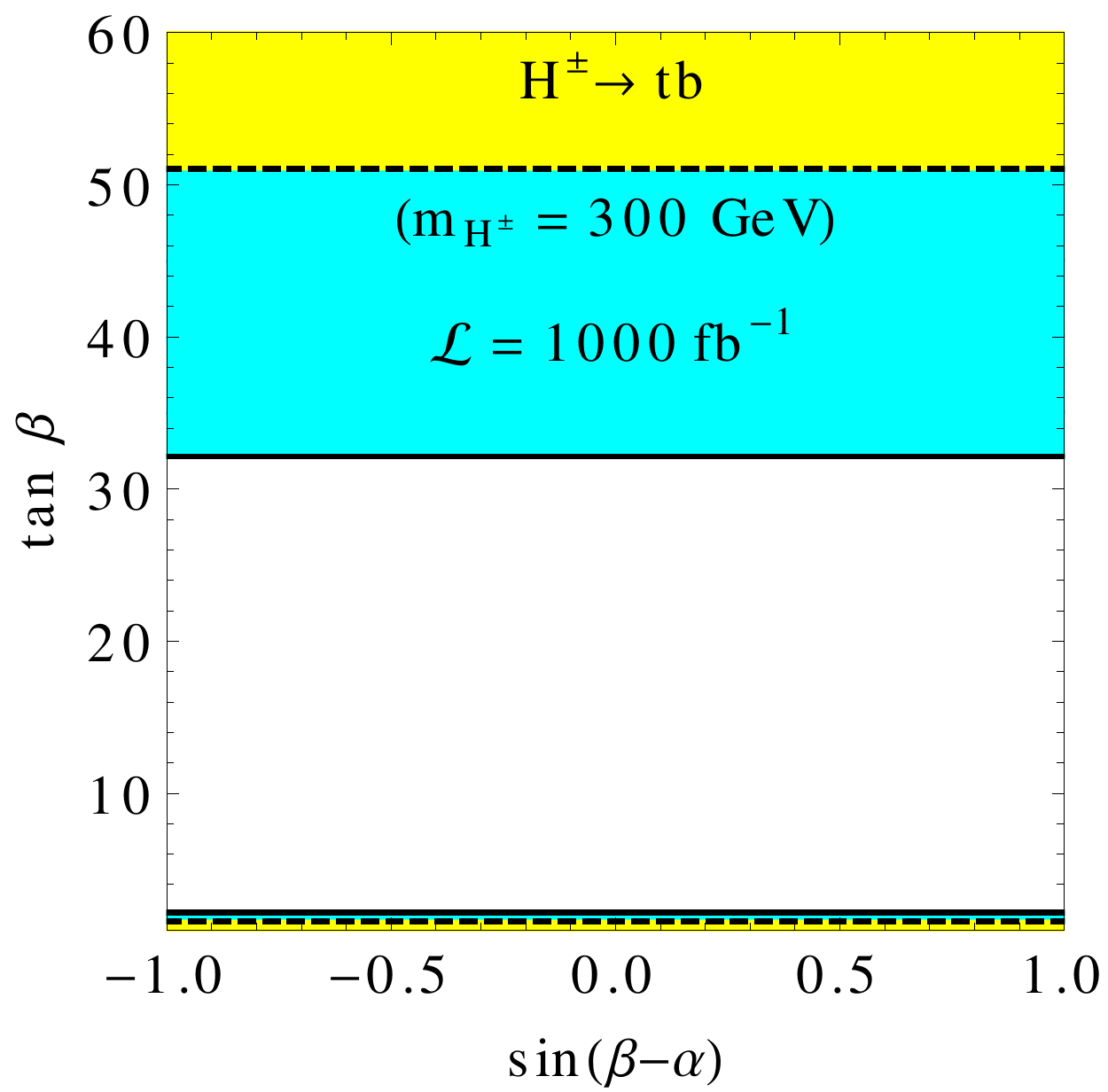}
\caption{The 95$\%$ exclusion (cyan regions) and the 5-$\sigma$ discovery reach(yellow regions) for the gaugophobic signal for $m_{H^{\pm}}$ 300 GeV. The integrated luminosity is fixed at $\mathcal{L}$ = 1000 fb$^{-1}$ at the 14 TeV LHC.} 
\label{fig:gaugo_contour}
\end{center}
\end{figure}

In Fig.~\ref{fig:gaugo_contour}, we show the contours for discovery and exclusion of a charged Higgs  in the $\tan\beta-\sin(\beta - \alpha)$ plane for the benchmark value $m_{H^{\pm}}$ = 300 GeV and for an integrated luminosity $\mathcal{L}$ = 1000 fb$^{-1}$. Bearing out the features of Fig.~\ref{fig:asso}, we see that the discoverable regions are close to $\tan\beta< 2$ $\tan\beta> 50$. The gaugophobic channel is independent of  $\sin(\beta - \alpha)$ -- we still choose to display the plot so as to be consistent across the different scenarios. From the  contour plot, one can see that 4$< \tan\beta <$35  is not optimal for charged Higgs discovery as the production cross-section is not sufficiently enhanced to overcome the SM background in this region. We find that, consistent with the current experimental results, the channel $gb\to\hpm t$ is not optimal for charged Higgs searches simply because of challenges related to the suppression of the SM background in this case.

\subsubsection{Chromophobic Models}
\begin{figure}[h!]
\includegraphics[scale=0.4]{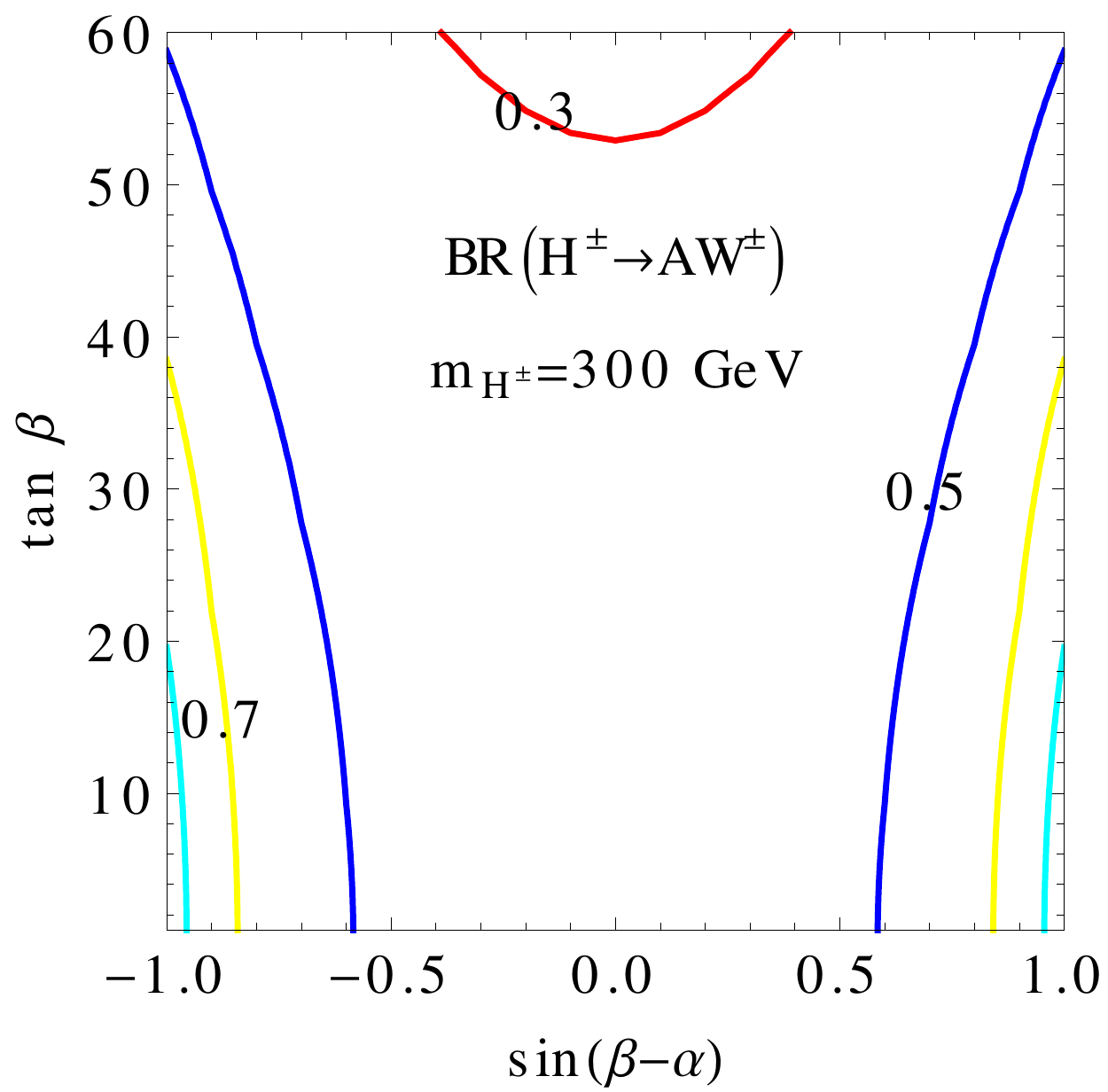}
\hspace{0.1in}
\includegraphics[scale=0.4]{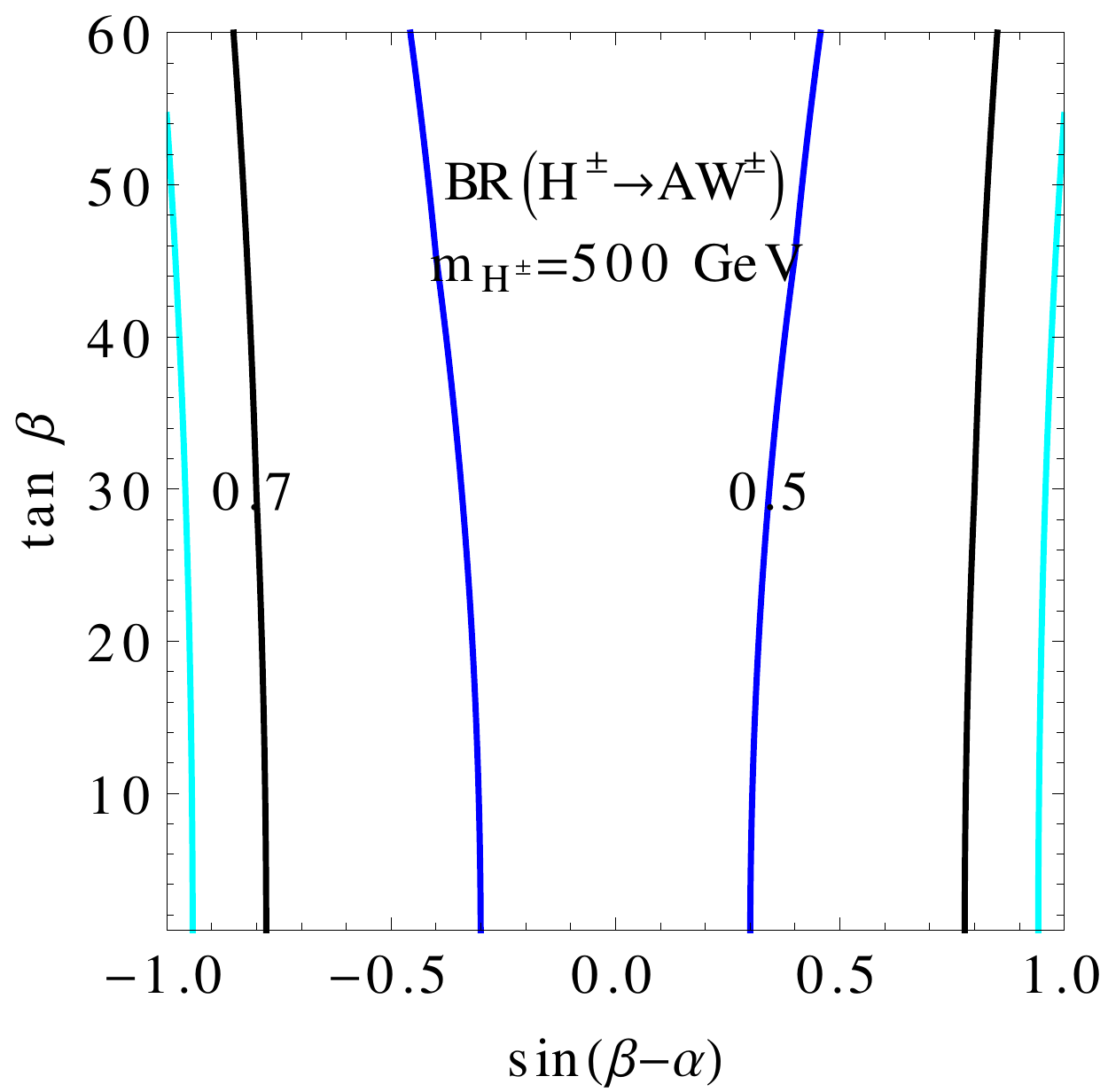}
\caption{The contour plot of $H^{\pm} \rightarrow W^{\pm} A$ channel for the parameter space $\sin(\beta - \alpha)$ versus $\tan\beta$ for $m_{H^{\pm}}$ = 300 GeV (left) and $m_{H^{\pm}}$ = 500 GeV (right).}
\label{fig:BRHcWA}
\end{figure}

In the case of the chromophobic signal, after production via the decay of a heavy scalar, the charged Higgs decays to a $W$ boson and a light scalar $A$. All couplings between $H^{\pm}$ and the colored particle are set to zero in keeping with the chromophobic nature of the charged Higgs. In Fig[\ref{fig:BRHcWA}] we present the contour plot of BR($H^{\pm} \rightarrow W^{\pm}A$) in the plane $\sin(\beta - \alpha)$ versus $\tan\beta$. From the Table[\ref{tab:classification}], one can see that the coupling $g_{H^{\pm}W^{\pm}A}$ does not have any dependence on $\alpha$ and $\beta$ -- thus the $\sin(\beta - \alpha)$ and $\tan\beta$ dependence arises due to the total width calculation where one needs to take into account all the available channels for the chromophobic charged Higgs. 

\begin{figure}[ht!]
\includegraphics[scale=0.4]{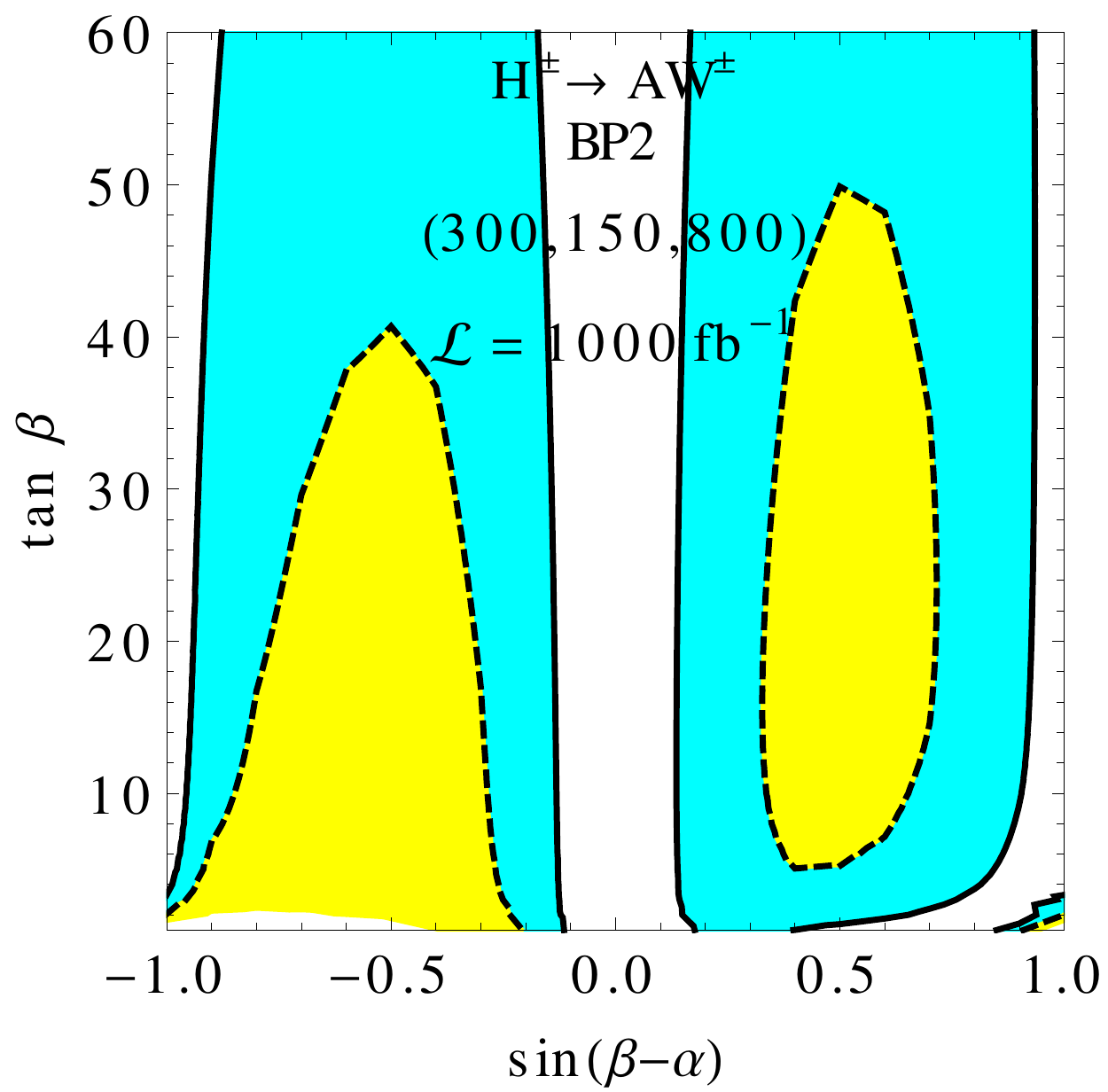}
\hspace{0.1in}
\includegraphics[scale=0.4]{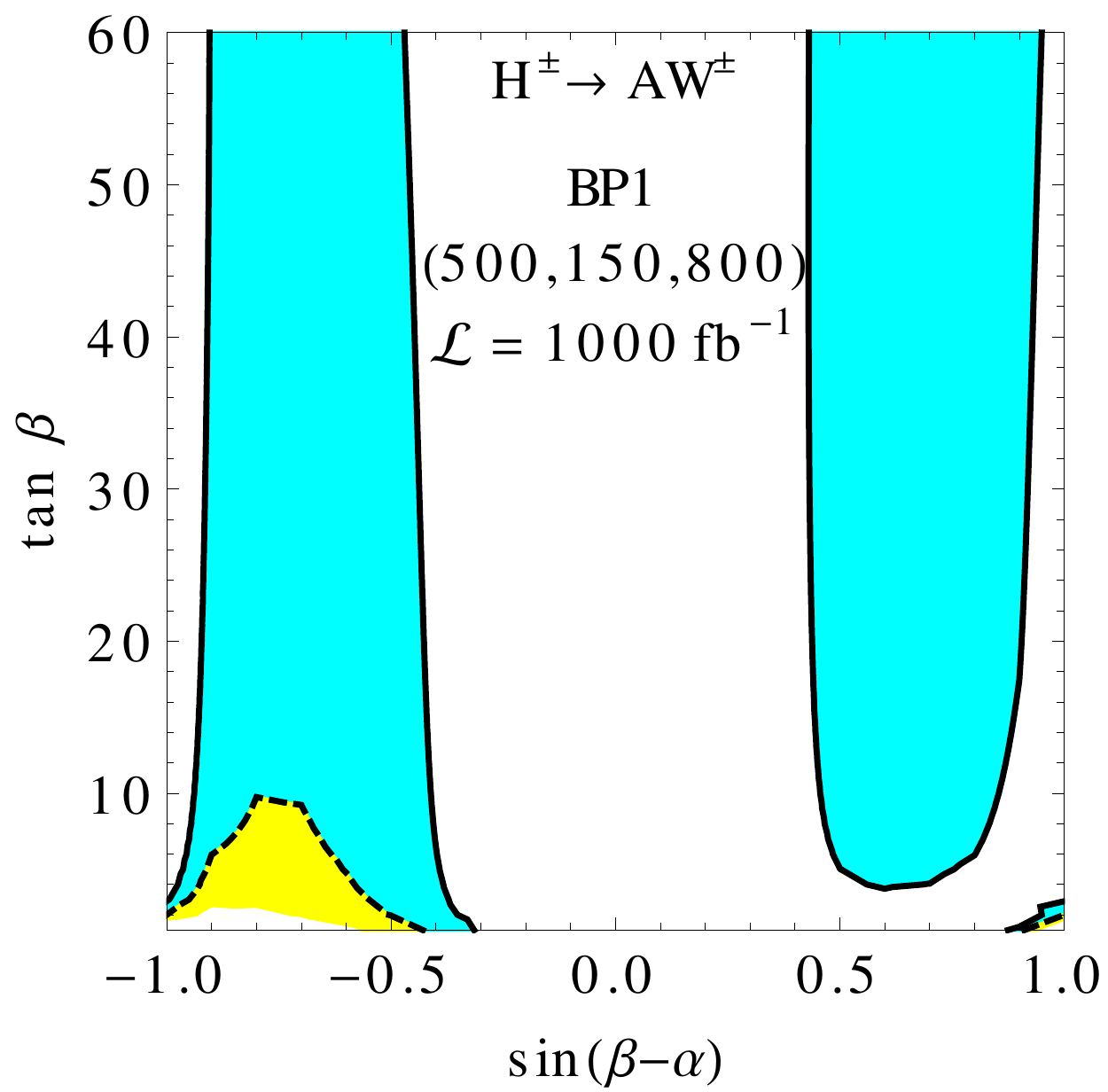}
\caption{The 95$\%$ exclusion regions(cyan regions) and the 5-$\sigma$ discovery reach(yellow regions) for the chromophobic signal in the $\sin(\beta - \alpha)$ verses $\tan\beta$ plane for the benchmark points 300 GeV (left) and 500 GeV (right).}
\label{fig:BR_chromo}
\end{figure}

One can observe from the plot that the BR becomes maximal in the $\sin(\beta - \alpha)\approx\pm1$ regions. Note that this is in contrast with the cross-section dependence on $\sin(\beta - \alpha)$ which becomes large in the complementary region  $\sin(\beta - \alpha)\approx\pm0$ (Fig~\ref{fig:Gluon}). Thus we expect the required $\sigma\times$BR for discovery/exclusion to happen for moderately large values of  $\sin(\beta - \alpha)$. In Fig[\ref{fig:BR_chromo}], we present the reach for the chromophobic signal in the $\sin(\beta - \alpha)$ versus $\tan\beta$ plane for the benchmark point 300 GeV, 500 GeV in which this feature is indeed borne out. The 300 GeV case has better reach in the parameter as compared to the 500 GeV due to the higher production cross-section rate. In this case, the $\hpm$ is discoverable in this channel in the regions $-1<\sin(\beta - \alpha)<-0.2$ and $\tan\beta<40$ and $0.35<\sin(\beta - \alpha)<0.5$ and $6<\tan\beta<50$. In the $m_{H^{\pm}}$ = 500 GeV case, the discovery region is confined to a small region $-1<\sin(\beta - \alpha)<-0.4$. Interestingly $\tan\beta$ = 7 is a potential discovery region for both benchmark values\footnote{If the charged Higgs is required to decay to a light CP-even Higgs, the discovery and exclusion regions are larger than that for the pseudoscalar case discussed here. However, such a light Higgs in 2HDM has to be 125 GeV which is not quite the benchmark point we have chosen (150 GeV).}.  

\subsubsection{Leptophobic Models}

\begin{figure}[h!]
\includegraphics[scale=0.4]{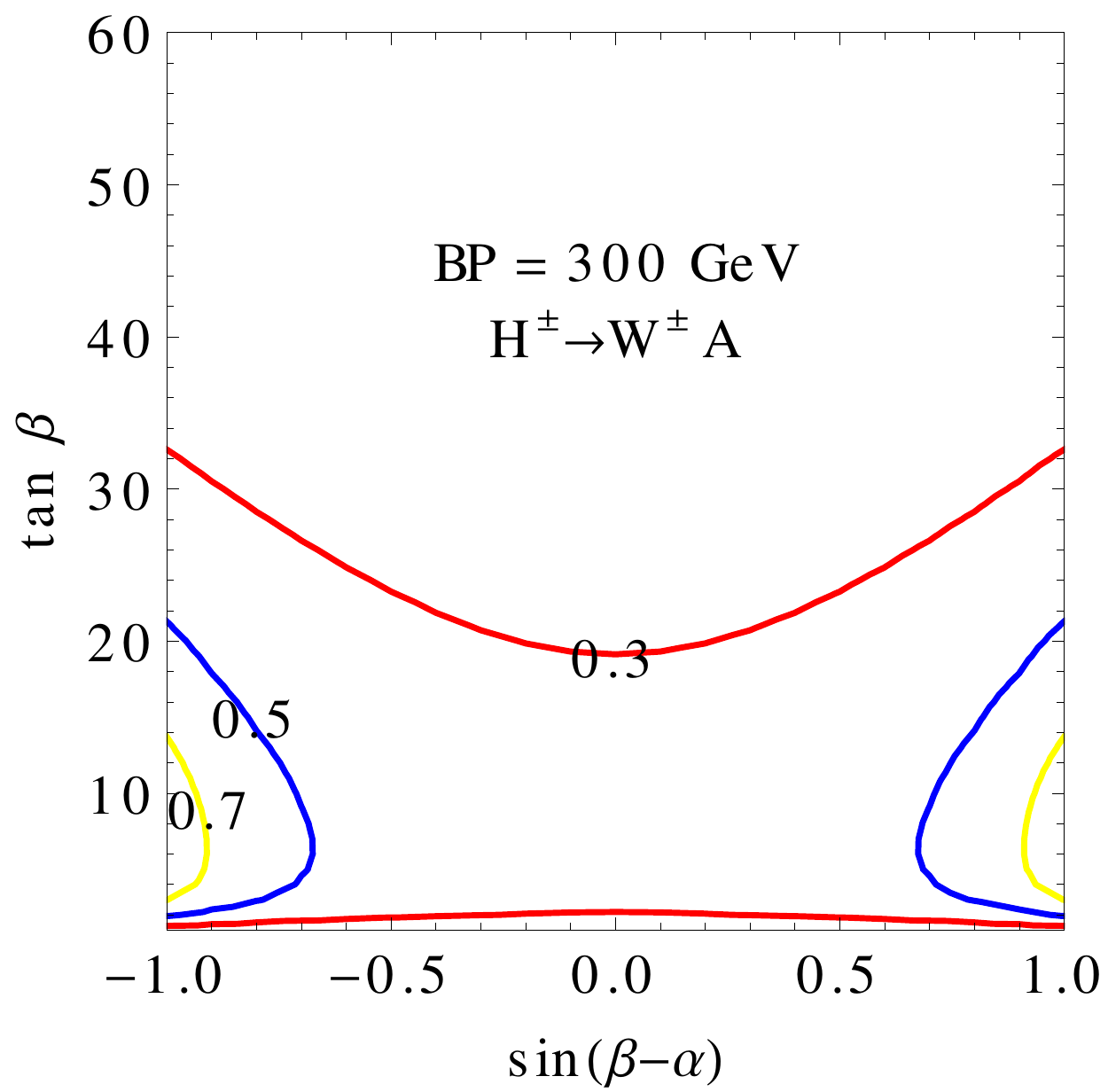}
\hspace{0.1in}
\includegraphics[scale=0.4]{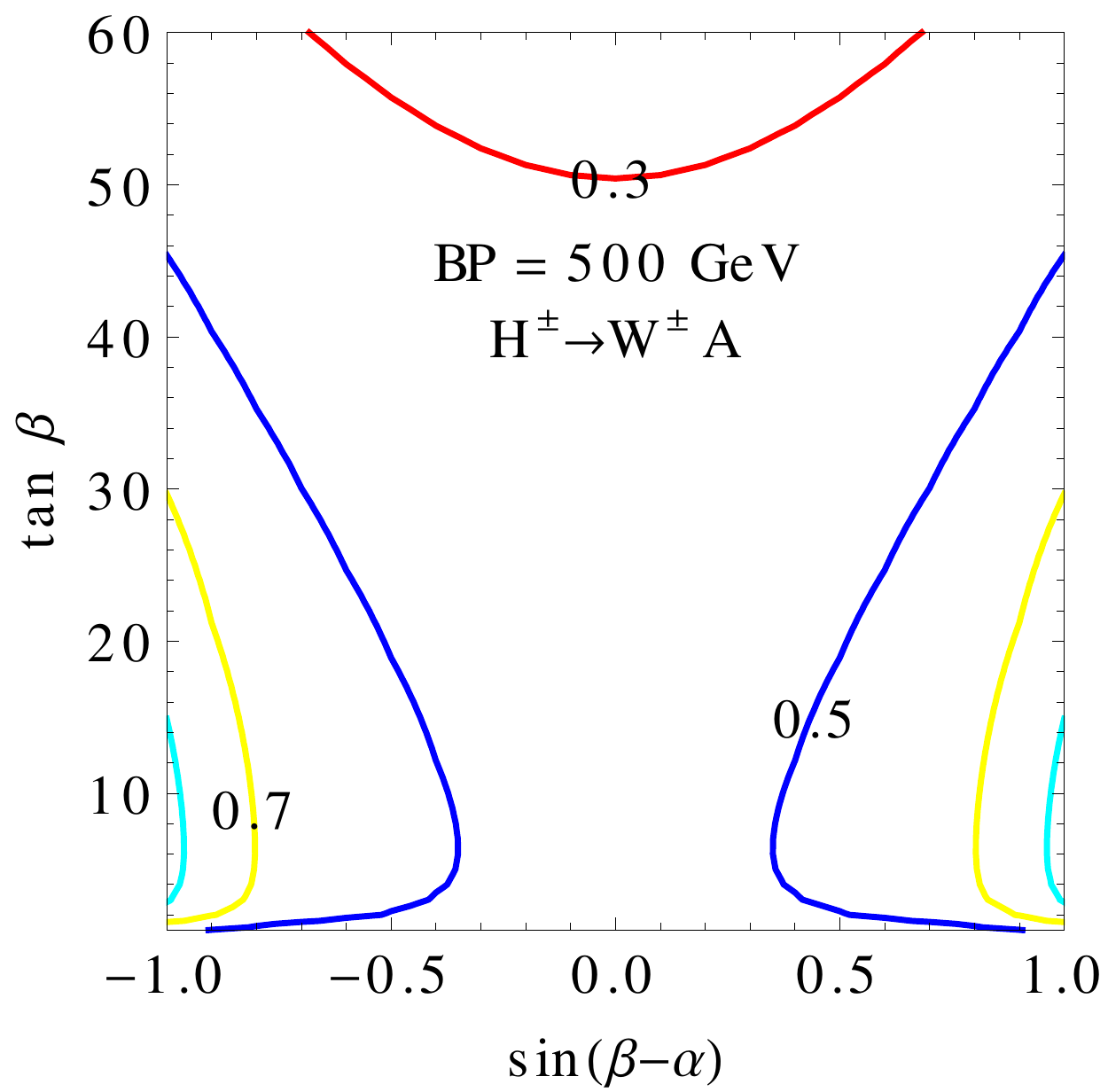}
\caption{The contour plot of BR($H^{\pm} \rightarrow W^{\pm} A$) in the $\sin(\beta - \alpha)-\tan\beta$ parameter space for $m_{H^{\pm}}$ = 300 GeV (left) and $m_{H^{\pm}}$ = 500 GeV (right).}
\label{fig:BRHcWA_Lep}
\end{figure}

\begin{figure}[ht!]
\includegraphics[scale=0.45]{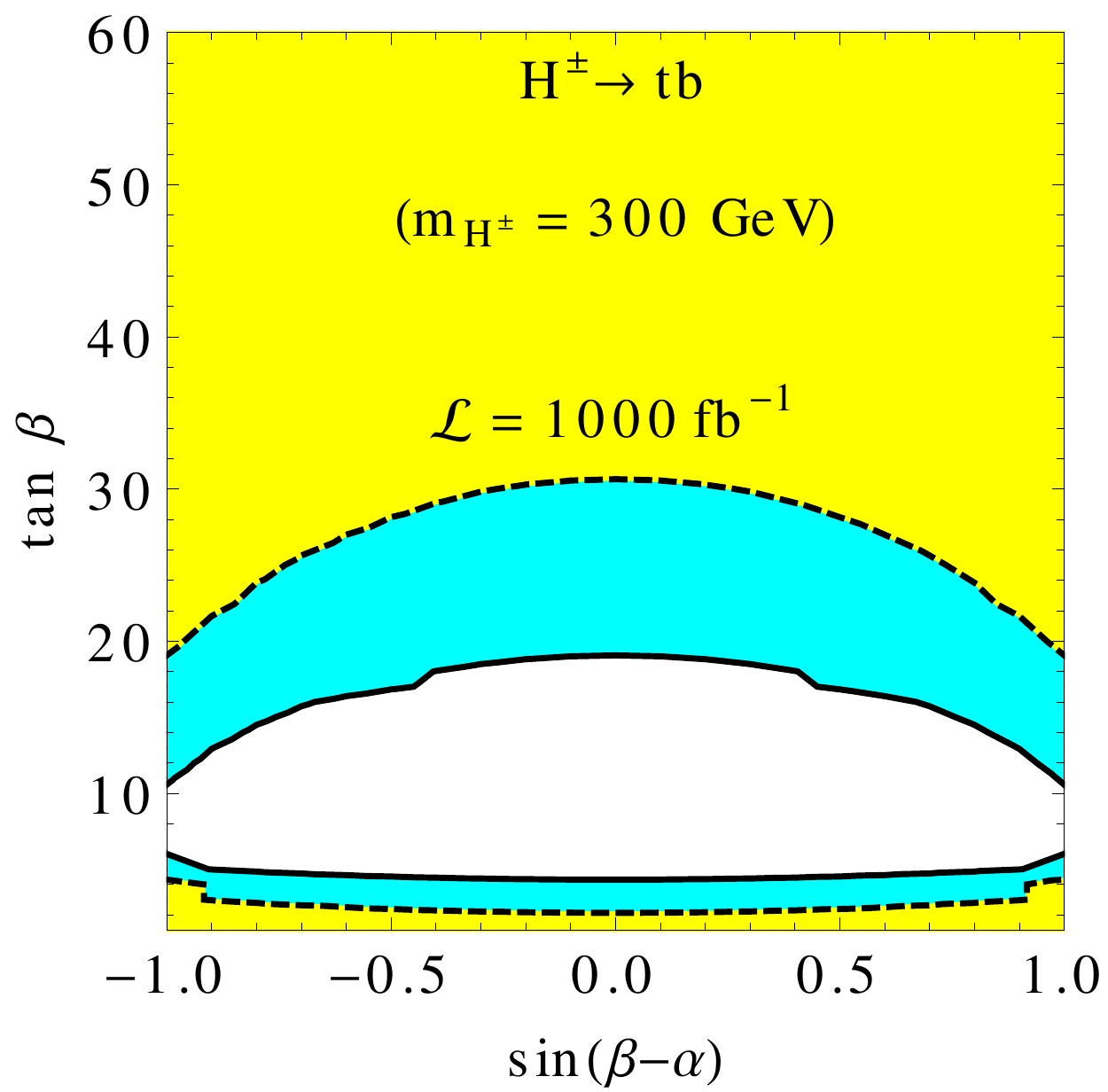}
\hspace{0.1in}
\includegraphics[scale=0.45]{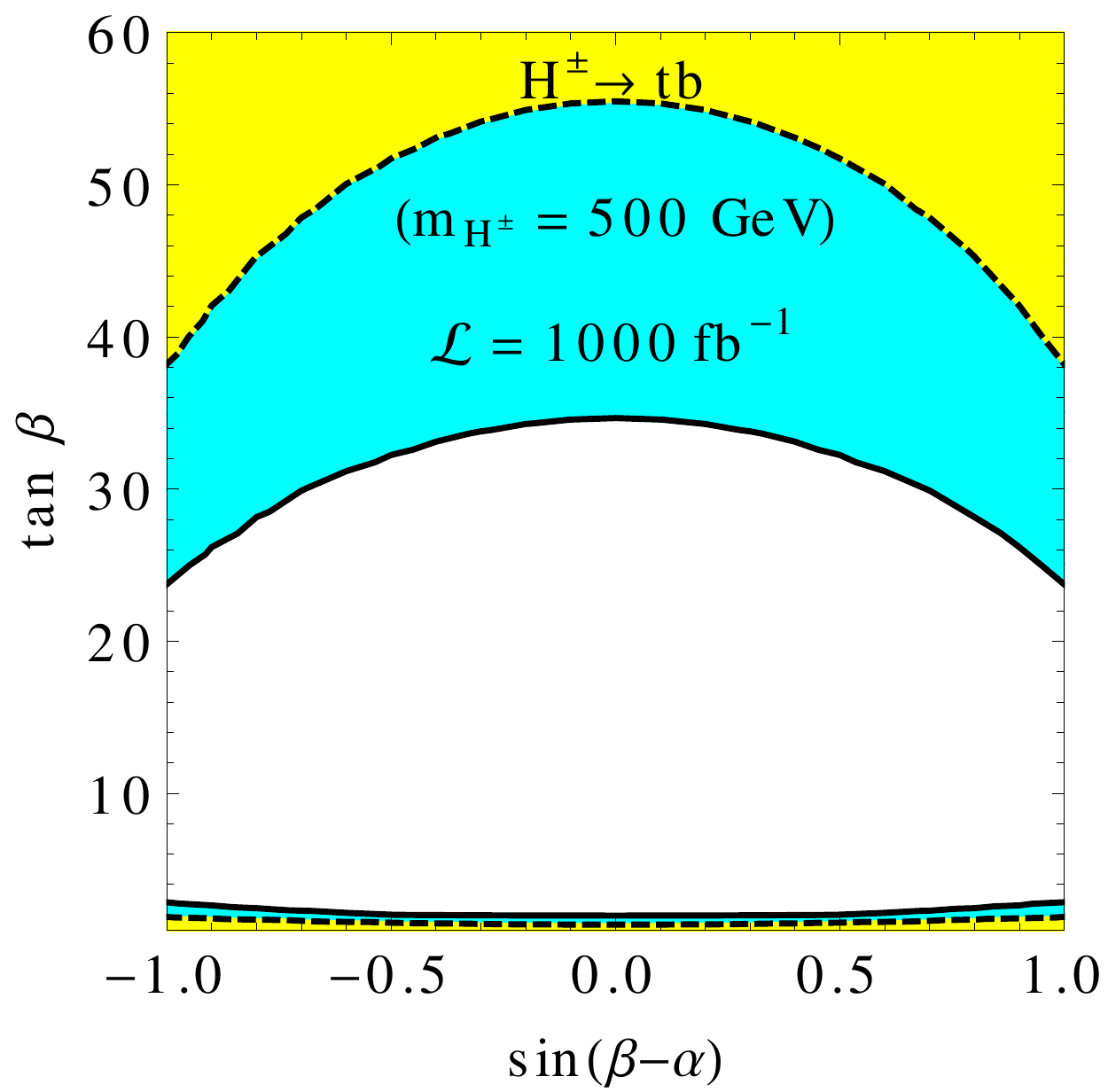}
\caption{The 95$\%$ exclusion regions(cyan regions) and the 5-$\sigma$ discovery reach(yellow regions) for the  leptophobic $3b+2j+\ell\nu$ signal for the benchmark points $m_{H^{\pm}}$ = 300 GeV (left) and $m_{H^{\pm}}$ = 500 GeV (right). It is seen that the entire range of $\sin(\beta - \alpha)$ for both small and large values of $\tan\beta$ are amenable to discovery.}
\label{fig:Lepto5_Contour}
\end{figure}

Leptophobic models can be probed in both the $2b+2j+\ell\nu$ and $3b+2j+\ell\nu$ channels. In the 5-jet process, the charged Higgs $H^{\pm}$ decays to $W$ boson and neutral Higgs $A$. Hence we need to calculate the branching ratio for $H^{\pm} \rightarrow W^{\pm}A$ setting the coupling between $H^{\pm}$ and $\tau \nu$ to zero. In Fig[\ref{fig:BRHcWA_Lep}] we present this branching ratio in the parameter plane $\sin(\beta - \alpha)$ versus $\tan\beta$. For the benchmark point $m_{H^{\pm}}$ = 300 GeV the branching ratio can be as high as 50$\%$  in the region $2 < \tan\beta < 20$ for $\sin(\beta - \alpha)\approx\pm 1$. Further, the BR is at least 30\% for the entire range of $\sin(\beta - \alpha)$ for moderately high $\tan\beta$. However, the overall numbers are not as high as in Fig~\ref{fig:BRHcWA} because the $\hpm\to tb$ (and the decay to other colored particles) takes up a significant amount of BR. However, it can be seen that in some regions of parameter space, this can be a viable decay channel to probe. The  $m_{H^{\pm}}$ = 500 GeV admits a better branching ratio as compared to the 300 GeV case as the decay is now more kinematically favored. In Fig~\ref{fig:Lepto5_Contour}], we present the the discovery and exclusion reach for this channel for an integrated luminosity of $\mathcal{L}$ = 1000 fb$^{-1}$. Both benchmark points have a potential 5$\sigma$ discovery reach complementary to the gaugophobic case (which only opened up for very high $\tan\beta$) and the chrompphobic case (which did not admit discovery in the region around $\sin(\beta - \alpha)\approx 0$).  While this channel offers the possibility of probing the charged Higgs for a wide range of $\sin(\beta - \alpha)$ values, its reach in terms of $\tan\beta$ is quite limited, again being restricted for small ($\lessapprox 4$) or large ($\gtrapprox 20$) values . The reach for the $m_{\hpm}=500$ GeV case is more restricted in spite of the larger branching ratios because the smaller production cross-section for the heavier charged Higgs is the decisive factor.

\begin{figure}[ht!]
\includegraphics[scale=0.4]{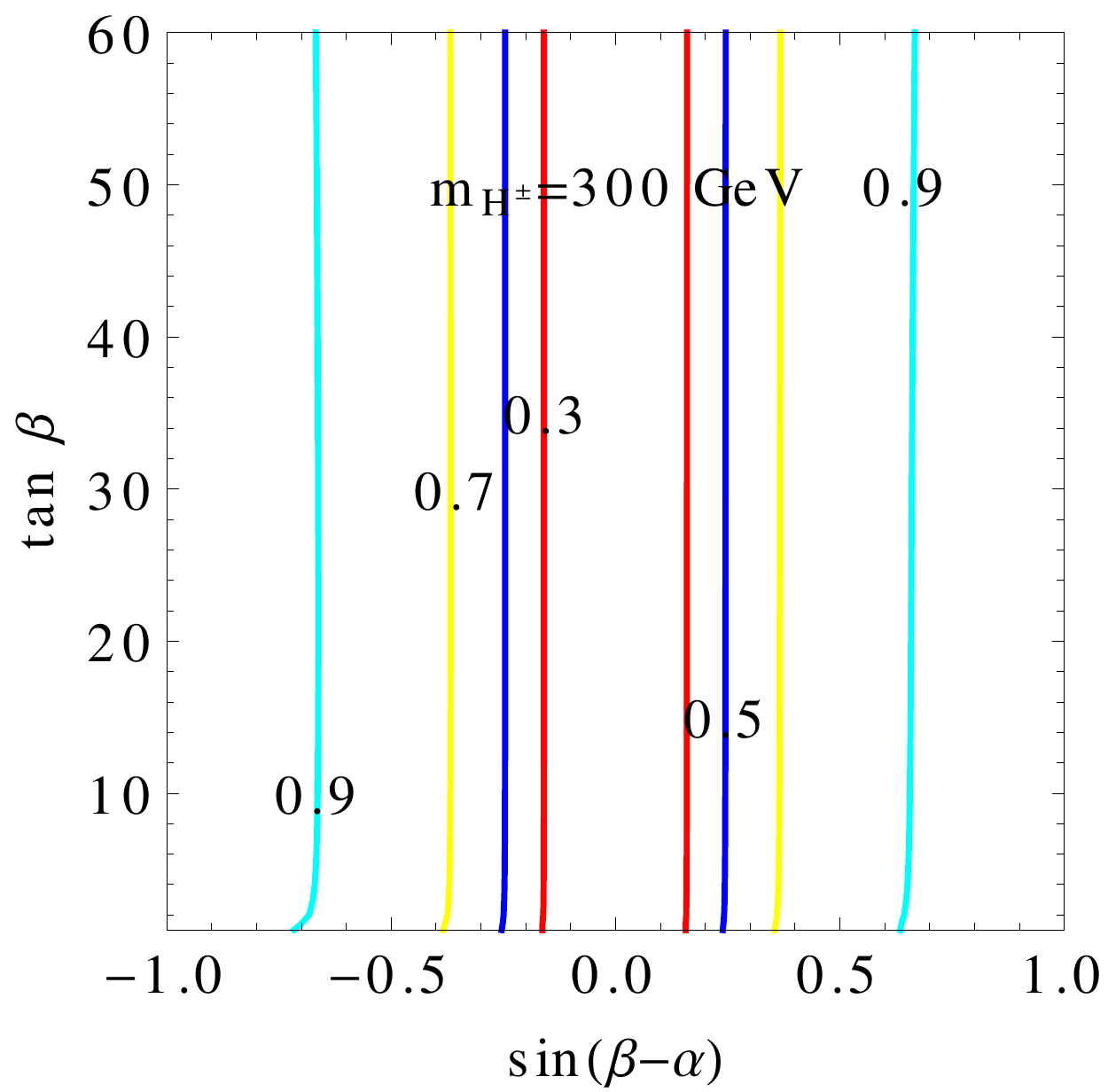}
\hspace{0.1in}
\includegraphics[scale=0.4]{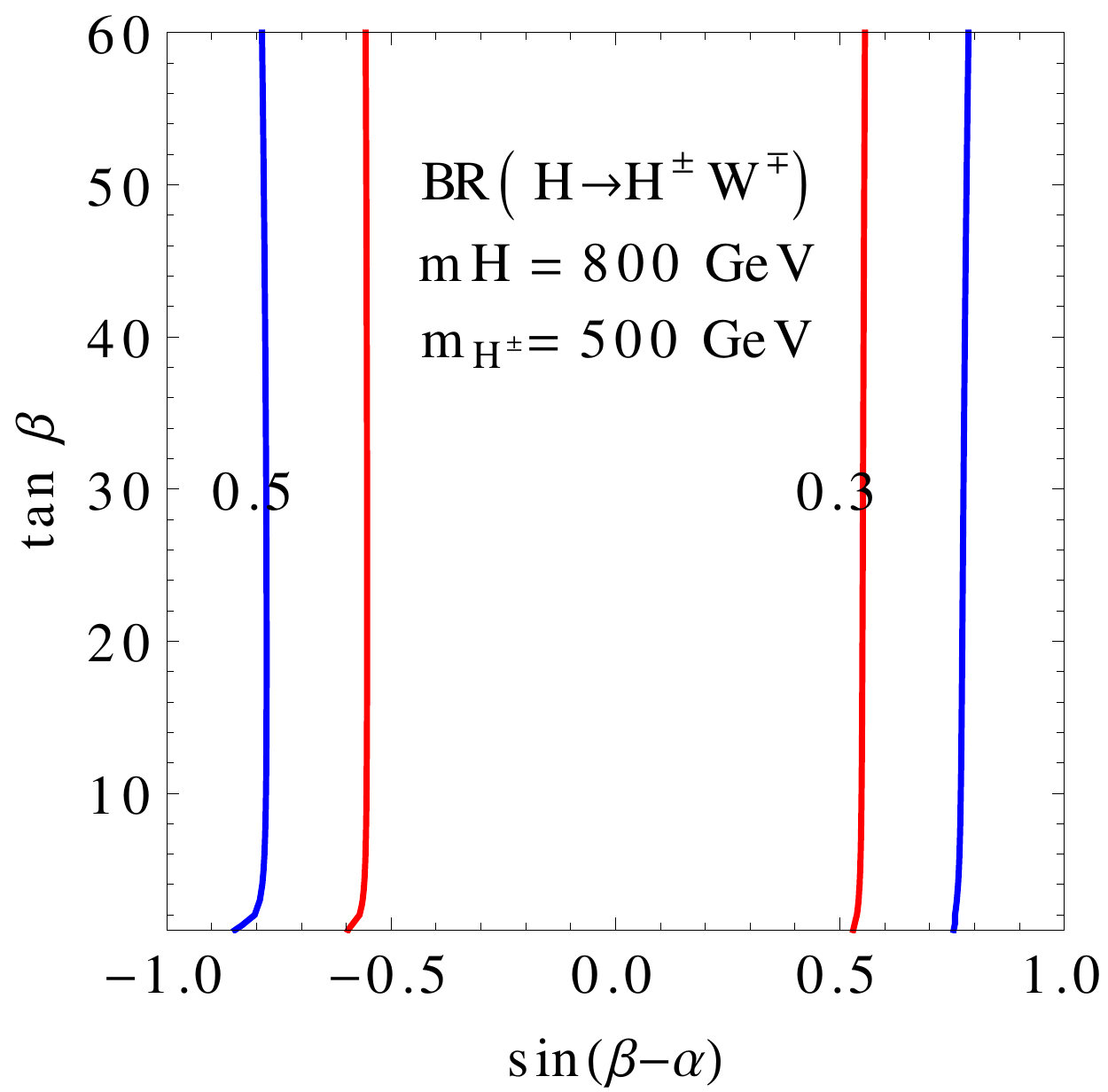}\\
\includegraphics[scale=0.4]{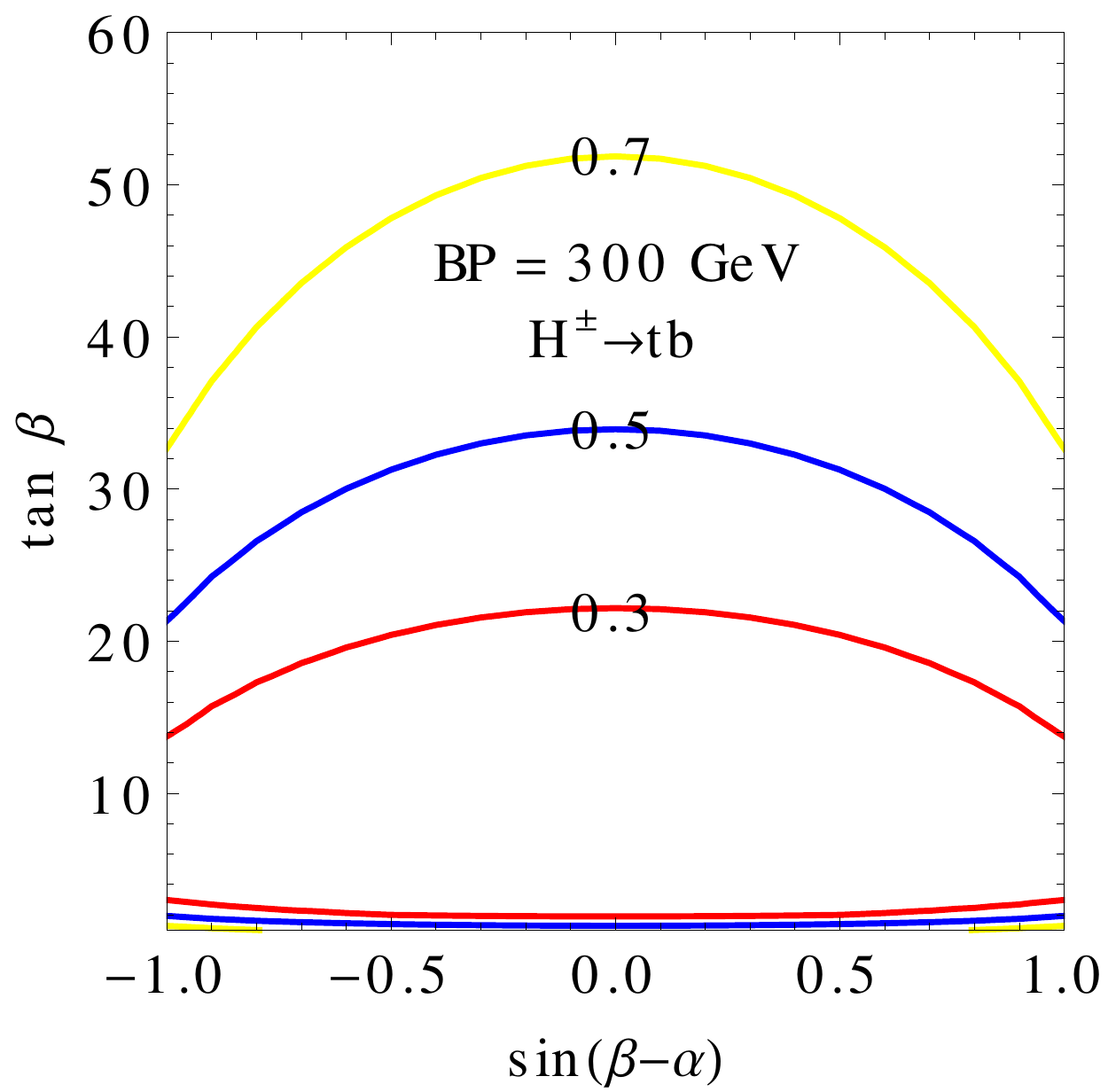}
\hspace{0.1in}
\includegraphics[scale=0.4]{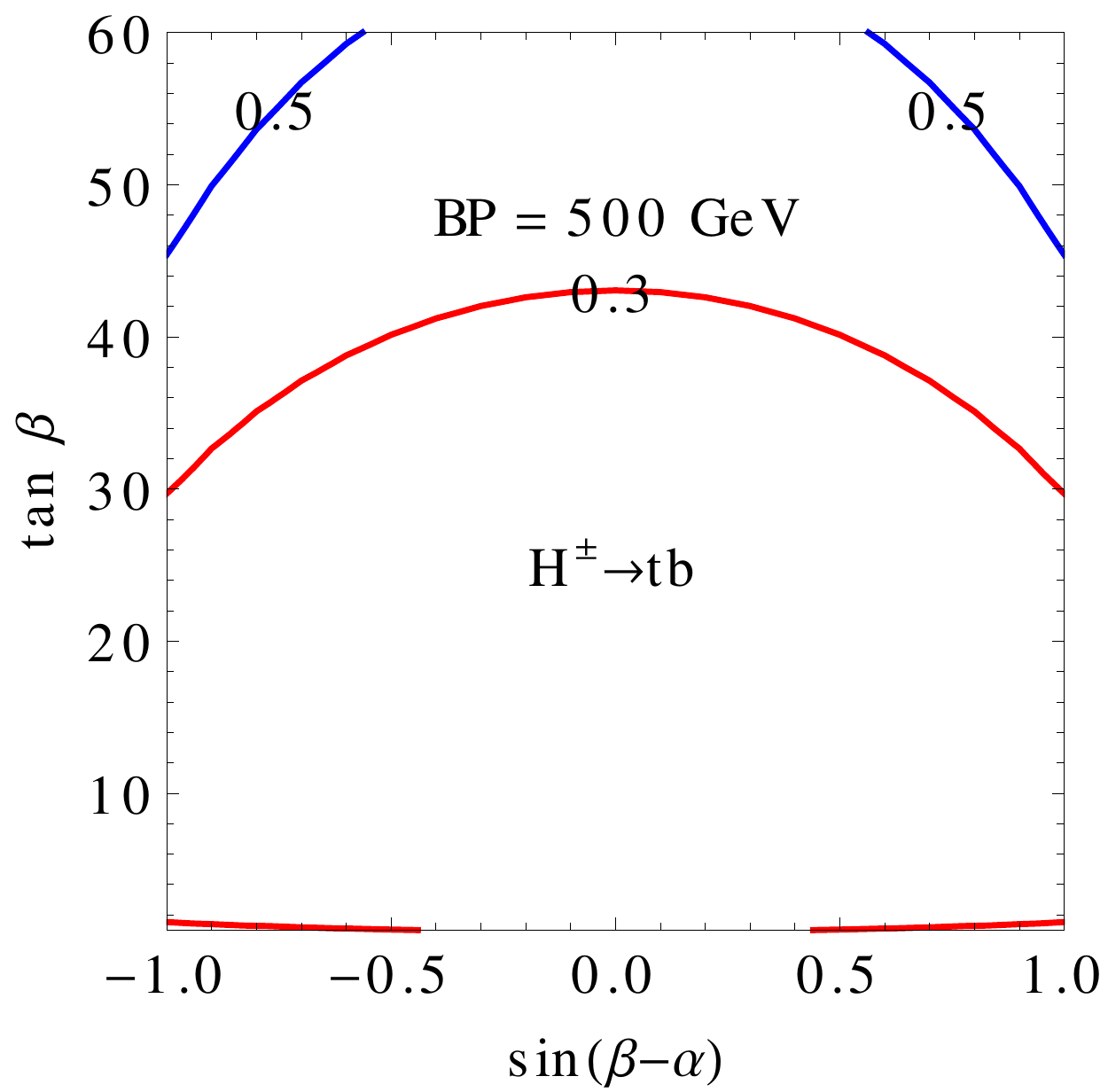}
\caption{(Top panel): Contours of the branching ratio of $H \rightarrow H^{\pm} W^{\pm}$ in the ($\sin(\beta -\alpha), \tan\beta$) plane for $m_H=$ 800 GeV and for $m_{H^{\pm}}=$ 300 GeV (left) and 500 GeV (right). (Bottom panel):  Contours of the branching ratio of $\hpm \rightarrow tb$ for $m_{H^{\pm}}=$ 300 GeV (left) and 500 GeV (right).} 
\label{fig:Lepto4-BR}
\end{figure}

In the Leptophobic-4-jets process, the charged Higgs decays is produced via the decay of a heavy scalar $H$ and further decays to $tb$ -- thus, in addition to the production cross-section of $H$, the relevant branching ratios for this process are those for $H\to \hpm \wmp$ and $\hpm\to tb$. In Fig~\ref{fig:Lepto4-BR}, we display contours of these two branching ratios in the ($\sin(\beta - \alpha)$,
$\tan\beta$) plane for $m_{\hpm}=300$ GeV and 500 GeV.  

\begin{figure}[ht!]
\begin{center}
\includegraphics[scale=0.4]{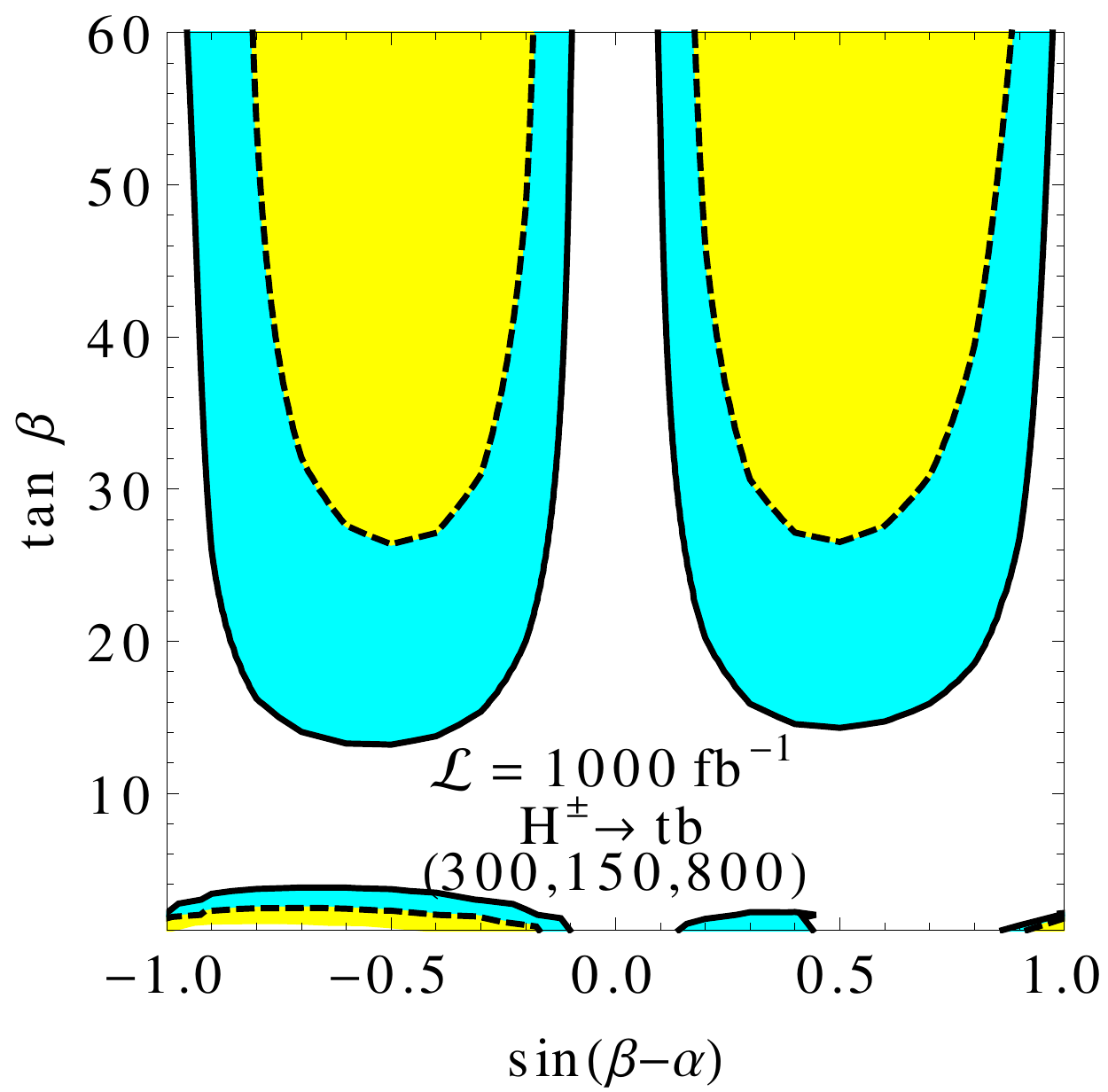}
\caption{The 95$\%$ exclusion regions(cyan regions) and the 5-$\sigma$ discovery reach(yellow regions) for the Leptophobic-4-jets signal in the $\sin(\beta - \alpha)-\tan\beta$ plane for the benchmark point $m_{H^{\pm}}=$ 300 GeV and assuming an integrated luminosity $\mathcal{L}$ = 1000 fb$^{-1}$ at the 14 TeV LHC. It is observed that the charged Higgs is discoverable in this process for $\tan\beta$ values greater than 30.}
\label{fig:Lepto4_Contour}
\end{center}
\end{figure}

It is seen that while BR$(H\to\hpm\wmp)$ is maximal towards $\sin(\beta -\alpha)=\pm 1$ for all values of $\tan\beta$, the BR$(\hpm\to tb)$ is appreciable for large and small values of $\tan\beta$. Remembering that the production cross-section of the $H$ (Fig.~\ref{fig:Gluon}) tends to favor moderately large $\sin(\beta -\alpha)$, we expect the product of these factors to be appreciable over a wide range of $\sin(\beta -\alpha)$ for both large and small values of $\tan\beta$. In Fig.~\ref{fig:Lepto4_Contour}, we have presented the discovery and exclusion contours for the charged Higgs in this channel for an integrated luminosity of $\mathcal{L}$ = 1000 fb$^{-1}$ -- it is seen that we indeed cover a wide range of parameter space. While the discovery regions for low $\tan\beta$ is rather limited confined to the region $-1\leq\sin(\beta -\alpha)\leq -0.2$, those for higher values of $\tan\beta$ are indeed appreciable. As opposed to the chromophobic case, the region $2<\tan\beta<10$ is immune to this search owing to the nature of the $g_{\hpm tb}$ coupling.

\section{Conclusions}
\label{sec:conclusions}

Numerous well motivated extensions of the SM incorporate an enlarged scalar sector with additional neutral and charged Higgs bosons. Now that the SM-like 125 GeV Higgs has been discovered in the ATLAS and CMS experiments, it behooves us to understand the potential of these experiments to unravel signatures of new physics. In this paper, we performed a complete collider analysis to understand the discovery potential of a charged Higgs boson in a model independent fashion by only assuming certainly general patterns in its coupling to the SM. In this spirit, we classified the charged Higgs to be gaugophobic, leptophobic, or chromophobic to understand the discovery reach of the charged Higgs in each case. Further, regardless of the pattern of the charged Higgs coupling, we were able to identify and classify the signals broadly in two categories: $2j+2b+\ell\nu$ and $2j+3b+\ell\nu$. Accordingly, the model independent part of the collider analysis dealt with devising effective cuts to suppress the backgrounds for both these processes from all SM sources ($t\bar{t}+$jets, $WZ+$jets) to make a 5$\sigma$ discovery of the $\hpm$ possible. 

Choosing benchmark points of $m_{\hpm}=$ 300 and 500 GeV, we find that the signal cross-sections required for the $\hpm$ discovery ranges between 6 and 13 fb for these different classes of charged Higgs couplings for an integrated luminosity of 500 fb$^{-1}$ and this range becomes 5-9 fb for $\mathcal{L}=$1000 fb$^{-1}$. We then proceeded to understand how viable such a scenario is from the point of view of a particular model -- which we chose to be the Type II 2HDM. We find that in the gaugophobic scenario, where the production and decay of the $\hpm$ is almost exclusively governed by the $g_{\hpm tb}$ coupling, one needs very low ($\leq$ 2) or high ($\geq$ 50) $\tan\beta$ for a 5$\sigma$ discovery of the charged Higgs, independent of the value of $\sin(\beta - \alpha)$ for $\mathcal{L}=$1000 fb$^{-1}$. In the chromophobic case, the efficacy of the analysis depends nontrivially on both $\sin(\beta - \alpha)$ and $\tan\beta$ and the discovery regions cluster around $-1<\sin(\beta - \alpha)<-0.2$ and $\tan\beta<40$ and $0.35<\sin(\beta - \alpha)<0.5$ and $6<\tan\beta<50$. This curious dependence on moderately large $\sin(\beta - \alpha)$ values is because while the production cross-section is enhanced in the small $\sin(\beta - \alpha)$ values, the relevant BR becomes appreciable only for larger values. Most importantly, this scenario admits discovery potential of the $\hpm$ in the region around $\tan\beta\approx$ 7 where traditional searches in the $\tau\nu$ final state typically are difficult. The leptophobic case, while displaying a different qualitative dependence of the discovery region on $\sin(\beta - \alpha)$ and $\tan\beta$ from the gaugophobic case, is similar to it in that one needs very low or very large $\tan\beta$ for discovery. The central point of the analysis is thus rather straightforward: if the charged Higgs couples to the colored sector of the SM, the dominant production and decay channels depend strongly on the $g_{\hpm tb}$ coupling and hence the features found in the gaugophobic and leptophobic scenarios emerge. If, however, one has an extension of the SM in which the charged Higgs does not couple to colored particles (i.e., to $tb$ in particular), one can have markedly different regions of the parameter space that become relevant for collider study.

We conclude this study by pointing out that it is imperative to probe for non-standard signatures of BSM physics in cases of extended scalar sectors. It is possible that depending on the nature of the charged Higgs couplings, the discovery of these particles can be effective in channels involving not one, but even two new physics couplings -- a case which is usually dismissed as non-viable might indeed turn out to be the dominant discovery mode. In fact, a discovery of the charged Higgs in one of these exotic channels might prove to be an efficient way of narrowing down the possibilities of new physics models at the TeV scale.

\begin{acknowledgments}
BC acknowledges support from the Department of Science and Technology, India, under Grant YSS/2015/001771. The work of SKR is partially supported by funding available from the Department of Atomic Energy, Government of India, for the Regional Centre for Accelerator-based ParticlePhysics (RECAPP), Harish-Chandra Research Institute. S.K. would like to thank the Theory division of CERN while the work was being completed. We thank Satendra Kumar for collaboration during the early stages of this project.
\end{acknowledgments}

\end{document}